\documentclass[prb,aps,twocolumn]{revtex4}
\usepackage{epsfig,epsf,psfrag}
\usepackage{graphicx}
\usepackage{epic,eepic}
\usepackage{color,pstcol}
\begin{document}
\title{\bf Photo-assisted Andreev reflection as a probe of quantum noise}
\date{\today}

\author{T. K. T. Nguyen$^{1,2,3}$, T. Jonckheere$^{1}$, A. Cr\'epieux$^{1,2}$, A. V. Nguyen$^{3}$, T. Martin$^{1,2}$}

\affiliation{$^1$ Centre de Physique Th\'eorique, Case 907 Luminy, 13288 Marseille cedex 9, France}
\affiliation{$^2$ Universit\'e de la M\'edit\'erann\'ee, 13288 Marseille Cedex 9, France}
\affiliation{$^3$ Institute of Physics and Electronics, 10 Dao Tan, Cong Vi, Ba Dinh, Hanoi, Vietnam}

\begin{abstract}
Andreev reflection, which corresponds to the tunneling of two
electrons from a metallic lead to a superconductor lead as a Cooper
pair (or vice versa), can be exploited to measure high frequency
noise. A detector is proposed, which consists of a normal
lead--superconductor circuit, which is capacitively coupled to a
mesoscopic circuit where noise is to be measured. We discuss two
detector circuits: a single normal metal -- superconductor tunnel
junction and a normal metal separated from a superconductor by a
quantum dot operating in the Coulomb blockade regime. A substantial
DC current flows in the detector circuit when an appropriate photon
is provided or absorbed by the mesoscopic circuit, which plays the
role of an environment for the junction to which it couples.
Results for the current can be cast in all cases in the form of a
frequency integral of the excess noise of the environment weighted
by a kernel which is specific to the transport process
(quasiparticle tunneling, Andreev reflection,...) which is
considered. We apply these ideas to the measurement of the excess
noise of a quantum point contact and we provide numerical estimates
of the detector current.
\end{abstract}
\maketitle

\section{Introduction}

Over the last decades, the measurement of noise has become a widely accepted
diagnosis in the study of electronic quantum transport\cite{yang, blanter_buttiker, chamon_freed96}.
Indeed, noise
provides information on the charge of the carriers in
unconventional conductors, when considering the Fano
factor -- the ratio of the noise to the average current --
in the regime where the carriers tunnel independently.
Away from this Poissonian regime, noise also contains
crucial information on the statistics of the charge
carriers \cite{martin_landauer}. Experimentally,
low frequency noise in the kHz--MHz range is more
accessible than high frequency (GHz--THz) noise: it can in principle
be measured using state of the art time
acquisition techniques. For higher frequency measurements,
it is becoming necessary to build a noise detector on chip
for a specific range of high frequencies.   In this work we
consider a detector circuit which is capacitively coupled to the
mesoscopic device. This circuit is composed of
a normal metal -- superconductor junction (NS junction).
Transport in this circuit occurs when the electrons tunneling between
the normal metal and the superconductor are either tunneling elastically,
or when they are able to gain or to loose
energy via photo-assisted Andreev reflections.
The ``photon'' is provided or absorbed from the
mesoscopic circuit which is capacitively coupled
to the NS detector.
The measurement of a DC current in the detector
can thus provide information on the
absorption and on the emission component of the current
noise correlator.

Several theoretical efforts have been made to describe
the high frequency noise measurement process \cite{lesovik_loosen}.
This is motivated by the fact that in specific transport setups, high
frequency noise detection is required in order to fully characterize
transport. Examples are Bell inequality tests
\cite{chtchelkatchev,lebedev_lesovik_blatter} and
the detection of the anomalous charges in one dimensional
correlated systems\cite{lebedev_crepieux_martin,dolcini}.
The former require information about high frequency noise
in order to establish a correspondence between electron number
correlators and current noise correlators. The latter requires
a high frequency noise measurement in order to obtain a
non-zero signal when the nanotube is connected to Fermi
liquid leads, which tend to wash out the features of a one
dimensional correlated electron system.

In Ref. \onlinecite{aguado_kouwenhoven}, a detection circuit,
which was capacitively coupled to the mesoscopic circuit
to be measured, was proposed as a high frequency noise
detector. This interesting theoretical idea has so far eluded
experimental verification, possibly because in the double dot system, additional (unwanted)
sources of inelastic scattering render a precise noise measurement quite
difficult. It is therefore necessary to look for detection circuits which
are less vulnerable to dissipation, as is the case of superconducting circuits,
because of the presence of the superconducting gap.
Indeed, the basic idea of Ref. \onlinecite{aguado_kouwenhoven} was implemented experimentally
recently\cite{deblock1,deblock2} using a
superconductor -- insulator -- superconductor (SIS) junction
as a detector, measuring in this case the finite frequency noise
characteristics of a Josephson junction.

More recently, the noise of a carbon nanotube/quantum dot
was also measured \cite{onac} using capacitive coupling to an SIS detector,
with the detection of Super-Poissonian noise
resulting from inelastic cotunneling processes.
The latter proposals, together with their successful experimental
implementations, indicate that superconducting detector circuits
have advantages over normal detection circuits because of the presence
of the superconducting gap.
\begin{figure}[t]
\centerline{\includegraphics[width=3.5cm,angle=90]{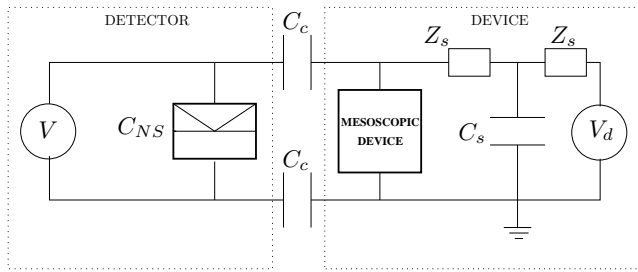}}
\caption{Schematic description of the set up: the mesoscopic device to be measured is coupled capacitively
to the detector circuit. The latter consists of a normal metal lead--superconductor junction with a DC bias.}
\label{figNS}
\end{figure}

The purpose of the present work is thus to analyze a similar situation,
except that the SIS junction is replaced by an NS circuit which
transfers two electrons using Andreev reflection between a normal lead and a superconductor.
The present scheme is similar in spirit to the initial proposal of Ref. \onlinecite{aguado_kouwenhoven},
in the sense that it exploits dynamical Coulomb blockade physics \cite{ingold_nazarov}.
However, here two electrons need to be transferred from / to the superconductor,
and such transitions involve high lying virtual states which are less prone to dissipation
because of the superconducting gap, similarly to the SIS detector of Refs. \onlinecite{deblock1,deblock2,onac}.
Andreev reflection\cite{btk} typically assumes a good contact
between a normal metal and a superconductor, but in general
it can be applied to tunneling contacts -- it then involves
tunneling transitions via virtual states.
Consequently, depending on the applied DC bias,
two successive ``inelastic'' electron
jumps are required for a current to pass through
the measurement circuit.
The amplitude of the DC current as a function
of bias voltage in the measurement circuit provides an effective readout
of the noise power to be measured.

\subsection{Detector consisting of a single NS junction}

The detector circuit is depicted in Fig. \ref{figNS}.
Two capacitors are placed, respectively,
between each side of the mesoscopic
device and each side of the NS
tunnel junction. This means that a current fluctuation in the mesoscopic
device generates, via the capacitors, a voltage fluctuation across the NS junction. In turn, the voltage fluctuations
translate into fluctuations of the phase around the junction.
The presence of the neighboring mesoscopic circuit acts as a
specific electromagnetic environment for this tunnel junction,
which is described in the context of dynamical Coulomb
blockade\cite{ingold_nazarov} for this reason.
\begin{figure}[h]

\centerline{\includegraphics[width=6.cm]{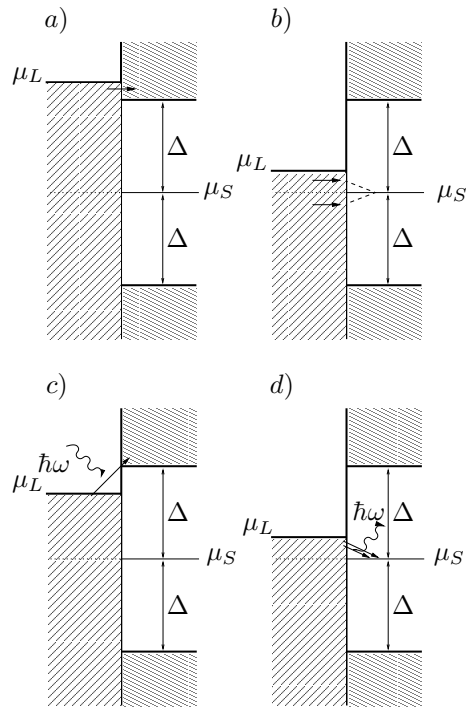}}
\caption{Electronic tunneling in a NS junction:
a) Quasiparticle electron tunneling. b) Andreev reflection. c) Photo-assisted electron tunneling as a quasiparticle in the superconductor. d) Photo-assisted Andreev reflection.
}
\label{NS_ener}
\end{figure}

Fig. \ref{NS_ener} depicts several scenarios for transport through
an NS interface. Elastic transfer of single electrons can occur if
the voltage applied to the junction is larger than the gap (Fig.
\ref{NS_ener}a). Below the gap, elastic transport can only occur via
Andreev reflection \cite{btk}, effectively transferring two electrons
with opposite energies with respect to the superconductor chemical
potential (Fig. \ref{NS_ener}b). Single electron can be transferred
with an initial energy below the gap, provided that a ``photon'' is
provided from the environment in order to create a quasiparticle in
the superconductor (Fig. \ref{NS_ener}c). Similarly, Andreev
reflection can be rendered inelastic by the environment: for
instance, two electrons on the normal side, with total energy above
the superconductor chemical potential, can give away a ``photon'' to
the environment, so that they can be absorbed as a Cooper pair in the
superconductor (Fig. \ref{NS_ener}d). As we shall see, such
inelastic Andreev processes are particularly useful for noise
detection.

\subsection{Detector consisting of an NS junction separated by a quantum dot}

\begin{figure}[h]
\centerline{\includegraphics[width=5.cm,angle=90]{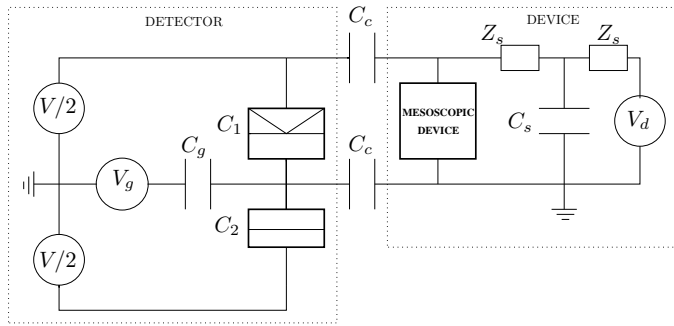}}
\caption{Schematic description of the NDS set up: the mesoscopic device to be measured is coupled capacitively
to the detector circuit. The latter consists of a normal metal lead -- quantum dot -- superconductor junction
with a DC bias.}
\label{figNDS}
\end{figure}

\begin{figure}[h]
\centerline{\includegraphics[width=6.cm]{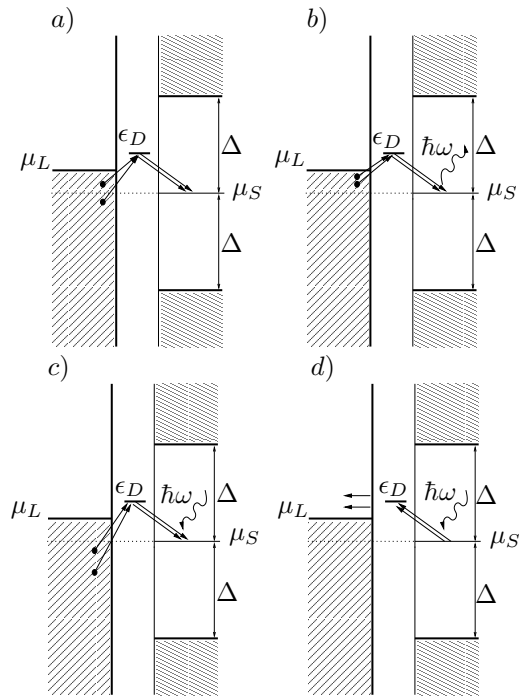}}
\caption{Andreev reflection in the NDS junction: a) Andreev reflection in the elastic regime. b), c) Photo-assisted Andreev reflection, where a ``photon''
is provided to or provided by the environment. d) Absorption of a
Cooper pair with (reverse) photo-assisted Andreev reflection, where a ``photon'' is provided by the environment.
For cases b), c), d), which require passing through the dot, the tunneling of electrons is sequential.
}
\label{NDS_ener}
\end{figure}

After studying the noise detection of the single NS junction, we
will turn later on to a double junction consisting of a normal metal lead, a quantum dot
operating in the Coulomb
blockade regime, and a superconductor connected to the latter (Fig. \ref{figNDS}).
The charging energy of the dot is assumed to be large enough
that double occupancy is prohibited.
This setup has the advantage on the previous proposal that additional
energy filtering is provided by the quantum dot.
Below, we refer to this system
as the normal metal--dot--superconductor (NDS) detector circuit.
In Fig. \ref{NDS_ener}, the quantum dot level is located
above the superconductor chemical potential, and placed well within the gap in 
order to avoid quasiparticle processes. Because double
occupancy is prohibited  by the Coulomb blockade, Andreev transport occurs
via sequential tunneling of the two electrons. Yet, because of energy
conservation, the same energy requirements as in Fig. \ref{NS_ener}b
have to be satisfied for the final states (electrons with opposite energies, see Fig. \ref{NDS_ener}a).
In T-matrix terminology, for this transition to occur, virtual states
corresponding to the energy of the dot are required, which suppress
the Andreev tunneling current because of large energy denominators
in the transition rate. Figs. \ref{NDS_ener} b, c, d, describe the cases where
an environment is coupled to the same NDS circuit. Provided that this
environment can yield or give some of its energy to the NDS detector, electronic
transitions via the dot can become much more likely because electron
energies on the normal side can be close to that of the dot level.
Such transitions can thus occur even if the chemical potential of the normal
lead exceeds that of the superconductor. As we shall see later on, the
bias voltage can act as a valve for photo-assisted electron transitions.
It is precisely these latter situations which will be exploited in order to measure the
noise of the measuring circuit (the ``environment'').

The paper is organized as follows. The model and the calculation of
the photo-assisted current in the NS junction are given in Sec. II,
and results for the detection of the noise of a quantum point
contact are presented. In Sec. III, we perform the same analysis for
the NDS junction and compare the results. The conclusion is point out in Sec. IV.

\section{Tunneling current through the NS junction}

\subsection{Model Hamiltonian}

The Hamiltonian which describes the decoupled normal metal lead--superconductor--environment
(mesoscopic circuit) system reads
\begin{equation}
H_0=H_{0_L}+H_{0_S}+H_{env}~,
\end{equation}
where
\begin{eqnarray}
H_{0_L}&=&\sum_{k,\sigma}\epsilon_k c^\dagger_{k,\sigma}c_{k,\sigma}~,
\end{eqnarray}
describes the energy states in the lead, with $c^\dagger_{k,\sigma}$ an electron creation operator.
The superconductor Hamiltonian has the diagonal form
\begin{equation}
H_{0_S}-\mu_S N_S=\sum_{q,\sigma}{E_q \gamma^\dagger_{q,\sigma}\gamma_{q,\sigma}}~,
\end{equation}
where $\gamma_{q,\sigma},\gamma_{q,\sigma}^{\dagger}$ are quasiparticle operators, which relate to the Fermi operators $c_{q,\sigma}, c^{+}_{q,\sigma}$ by the Bogoliubov transformation
\begin{eqnarray}
c_{-q,\downarrow}&=&u_q \gamma_{-q,\downarrow}-v_q \gamma^\dagger_{q,\uparrow}\nonumber~,\\
c^\dagger_{q,\uparrow}&=&u_q \gamma^\dagger_{q,\uparrow}+v_q \gamma_{-q,\downarrow}~,
\end{eqnarray}
and $E_q=\sqrt{\Delta^2+\zeta^{2}_{q}}$ is the quasiparticle energy, $\zeta_q=\epsilon_q-\mu_S$ is the normal state single-electron energy counted from the Fermi level $\mu_S$, $\Delta$ is the superconducting gap
which will be assumed to be the largest energy scale in these calculations. Hereafter, we also define $eV=\mu_L-\mu_S$ and assume $\mu_S=0$.

Here we do not specify the Hamiltonian of the environment because the environment represents an open system:
the mesoscopic circuit which represents the environment will only
manifest itself via the phase fluctuations $\langle \phi (0)\phi(t)\rangle$ which are induced at the NS junction (or later on for the
NDS circuit, at the dot--superconductor junction) because of the location of the capacitor plates.
In what follows, we shall assume that the unsymmetrized noise spectral density
\begin{equation}
S^+(\omega)=2\int_{-\infty}^{+\infty} dt e^{i\omega t}
    \langle\langle I(0) I(t)\rangle\rangle~,
\label{C+}
\end{equation}
corresponding to photon emission (for positive frequency), or alternatively $S^-(\omega)\equiv S^+(-\omega)$,
the spectral density of noise corresponding to photon absorption, are both specified by the
transport properties of the mesoscopic circuit\cite{lesovik_loosen,gavish_levinson_imry,martin_houches}.
Here $\langle\langle\ldots\rangle\rangle$
stands for an irreducible noise correlator, where the product of average currents has been subtracted out.

The tunneling Hamiltonian describing the electron transferring between the superconductor and normal metal lead in the NS junction is
\begin{equation}
H_T=\sum_{k,q,\sigma}T_{k,q} c^\dagger_{k,\sigma}c_{q,\sigma}e^{-i\phi}~,
\end{equation}
the indices $k$, $q$ refer to the normal metal lead, superconductor.
 We consider for simplicity that $T_{k,q}=T_0$.
Note that the tunneling Hamiltonian contains a fluctuating phase factor, which represents the coupling to the mesoscopic circuit. Indeed, because of the capacitive coupling between the sides of the NS junction
and the mesoscopic circuit, a current fluctuation translates into a voltage fluctuation across the NS junction. Both are related via the trans-impedance of the circuit \cite{aguado_kouwenhoven} $ V(\omega)=Z(\omega)I(\omega)$.
Next, the voltage fluctuations translate into phase fluctuations across the junction, as the phase
is the canonical conjugate of the charge at the junctions\cite{ingold_nazarov}: the phase is thus
considered as a quantum mechanical operator throughout this paper.
For definition purposes, it is convenient to introduce
the correlation of the phase operators
\begin{equation}
J(t)=\langle[\phi(t)-\phi(0)]\phi(0)\rangle~,
\label{jt}
\end{equation}
where the phase operator is related to the voltage by the
relation
\begin{equation}
\phi(t)=e\int_{-\infty}^{t}\!\!\! dt^\prime V(t^\prime)~.
\end{equation}
Given a specific circuit (capacitors, resistances,...) 
the phase correlator is therefore expressed in terms of the trans-impedance of the circuit and the
spectral density of noise\cite{aguado_kouwenhoven}
\begin{equation}
J(t)=\frac{2\pi}{R_K}\int_{-\infty}^{\infty}\!\!\!\!\! d\omega\frac{|Z(\omega)|^2}{\omega^2}S_I(\omega)(e^{-i\omega t}-1)~,
\end{equation}
where $R_K=2\pi/e^2$ is the quantum of resistance and $S_{I}(\omega)=S^{+}(-\omega)$.

The present system bears similarities with the study of inelastic Andreev reflection in the case where
the superconductor contains phase fluctuations\cite{falci,devillard}. Such phase fluctuations destroy
the symmetry between electrons and holes, and affect the current voltage characteristics of the NS
junction.

\subsection{Tunneling current}

The tunneling current associated with two electrons is given by the Fermi Golden Rule $I=2e\Gamma_{i\rightarrow f}$, with the tunneling rate\cite{notice}
\begin{equation}
\Gamma_{i\rightarrow f}=2\pi\sum_f {|\langle f| T |i \rangle|}^2 \delta(\epsilon_i-\epsilon_f)~,
\label{rate}
\end{equation}
where $\epsilon_i$ and $\epsilon_f$ are the tunneling energies of the inital and final states, including the environment,
$T$ is the transition operator, which is expressed as
\begin{eqnarray}
T&=& H_T+H_T\sum_{n=1}^{\infty}\left(\frac{1}{i\eta-H_0+\epsilon_i}H_T\right)^n~,
\end{eqnarray}
with $\eta$ is a positive infinitesimal.

Throughout this paper, one considers the photo-assisted tunneling (PAT) current due to the high frequency current
fluctuations of the mesoscopic device, as the difference \cite{ingold_nazarov}:
\begin{equation}
I_{PAT}=I(\text{environment})-I(\text{no environment})~,
\end{equation}
where, in general, the total current for tunneling of electrons through the junction is $I=I_\rightarrow-I_\leftarrow$.

However, experimentally\cite{deblock2} it is difficult to couple capacitively and then to remove the mesoscopic device circuit from the detector circuit. What is in fact often measured is the excess noise, i.e. the difference between current fluctuations at a given bias and zero bias in the mesoscopic circuit. In the latter on, we will calculate the excess noise $S^+_{ex}(\omega)$ of $S^+(\omega)$.
In this work, we thus measure the difference between the currents through the detector when the device circuit is applied a bias voltage and zero bias, as a function of detector bias voltage, which is defined as
\begin{equation}
\Delta I_{PAT}(eV)=I(eV_d\neq 0,eV)-I(eV_d=0,eV)~.
\end{equation}
This also corresponds to the difference between the PAT currents through the detector when the device circuit is applied a bias voltage and zero bias $\Delta I_{PAT}(eV)=I_{PAT}(eV_d\neq 0,eV)-I_{PAT}(eV_d=0,eV)$ because the contributions
with no environment cancel out.
The difference PAT current thus provides crucial information on the spectral density of
excess noise of the mesoscopic device. Notice that our calculation applies to the
zero temperature case for convenience, but it can be generalized to
finite temperatures.

\subsection{Single electron tunneling}

Although our focus of interest concerns photo-assisted Andreev
reflection, we need to compute all possible contributions. The
current associated with one electron tunneling is given by the Fermi
golden rule:
\begin{equation}
I=2\pi e\sum_f {|\langle f| H_T |i \rangle|}^2 \delta(\epsilon_i-\epsilon_f)~,
\end{equation}

The calculation of the current proceeds in the same way as that of a normal metal junction
\cite{ingold_nazarov}, except that one has to take into account the superconducting
density of states on the right side of the junction which is done by exploiting the Bogoliubov
transformation.
For the case of electrons tunneling from the normal metal lead to the superconductor $(eV>\Delta)$, the current from left to right reads
\begin{widetext}
\begin{eqnarray}
I_\rightarrow&=&e\int_{-\infty}^{\infty}\!\!\!dt e^{-i(\mu_S-\mu_L)t}\langle H_T(t) H_T^\dagger(0)\rangle\nonumber\\
&=& 4\pi eT_0^2\mathcal N_N^2 \int_{-\infty}^{eV}d\epsilon\int_{\Delta}^{\infty}dE\frac{E}{\sqrt{E^2-\Delta^2}}\left[1-\frac{2\pi}{R_K}\int_{-\infty}^{\infty}d\omega\frac{|Z(\omega)|^2}{\omega^2}S_I(\omega)\right]\delta(E-\epsilon)\nonumber\\
 &&+\frac{8\pi^2 eT_0^2\mathcal N_N^2}{R_K} \int_{-\infty}^{eV}d\epsilon\int_{\Delta}^{\infty}dE\frac{E}{\sqrt{E^2-\Delta^2}}\frac{|Z(\epsilon-E)|^2}{(\epsilon-E)^2}S_I(\epsilon-E)~,
\end{eqnarray}
\end{widetext}
with $R_K$ the quantum of resistance.
This current includes both an elastic and an inelastic contribution, the former being renormalized by the presence
of the environment. Here, $\epsilon$ is the energy of an
electron in the normal metal lead and $E$ is the energy of a quasiparticle in the Superconductor lead.
Changing variables in the inelastic term to $\Omega=\epsilon-E$, $\delta=\epsilon+E$,
an using $\int dx (x+a)/\sqrt{(x+a)^2-b^2}=\sqrt{(x+a)^2-b^2}/2$, after computing the current from right to left in
a similar way, we obtain
\begin{eqnarray}
\Delta I_{PAT}&=&-C_{1e}\left(\int_{-\infty}^{\infty}d\omega\frac{|Z(\omega)|^2}{\omega^2}S^+_{ex}(-\omega)\right)K^{el}_{1e}(eV)\nonumber\\
&&\!\!\!\!\!\!\!\!\!\!\!\!\!\!\!\!\!\!\!\!\!\!+C_{1e}\int_{-\infty}^{eV-\Delta}d\Omega\frac{|Z(\Omega)|^2}{\Omega^2}S^+_{ex}(-\Omega)K^{inel}_{1e}(\Omega,eV)~,
\end{eqnarray}
where the transmission coefficient of the NS junction in the normal state is defined as $\mathcal{T}=4\pi^2\mathcal{N}_{L}\mathcal{N}_{0_S}T_0^2$, $C_{1e}=e\mathcal{T}/R_K$. The weight functions are defined as
\begin{equation}
K^{el}_{1e}(eV)=\sqrt{(eV)^2-\Delta^2}~,
\end{equation}
\begin{equation}
K^{inel}_{1e}(\Omega,eV)=\sqrt{(\Omega-eV)^2-\Delta^2}~.
\end{equation}

Similarly, we obtain the formula of $\Delta I_{PAT}$ for the case $eV\leq-\Delta$
\begin{eqnarray}
\Delta I_{PAT}&=&-C_{1e}\left(\int_{-\infty}^{\infty}d\omega\frac{|Z(\omega)|^2}{\omega^2}S^+_{ex}(-\omega)\right)K^{el}_{1e}(eV)\nonumber\\
&&\!\!\!\!\!\!\!\!\!\!\!\!\!\!\!\!\!\!\!\!\!\!+C_{1e}\int^{\infty}_{eV+\Delta}d\Omega\frac{|Z(-\Omega)|^2}{\Omega^2}S^+_{ex}(\Omega)K^{inel}_{1e}(\Omega,eV)~.
\end{eqnarray}
For the case $-\Delta\leq eV\leq\Delta$: there are no elastic transitions because electrons cannot enter the superconducting gap.
Nevertheless, due to the presence of the environment, electrons can absorb/emit energy from/to the environment so that
an inelastic quasiparticle current flows.
\begin{eqnarray}
\Delta I_{PAT}
&=&C_{1e}\int_{-\infty}^{eV-\Delta}d\Omega\frac{|Z(\Omega)|^2}{\Omega^2}S^+_{ex}(-\Omega)K^{inel}_{1e}(\Omega,eV)\nonumber\\
&&\!\!\!\!\!\!\!\!\!\!\!\!\!\!\!\!\!\!\!\!\!\!-C_{1e}\int^{\infty}_{eV+\Delta}d\Omega\frac{|Z(-\Omega)|^2}{\Omega^2}S^+_{ex}(\Omega)K^{inel}_{1e}(\Omega,eV)~.
\end{eqnarray}

\subsection{Two electrons tunneling as two quasiparticles}

The single electron (quasiparticle) tunneling current is of first order in the tunneling amplitude.
We now turn to processes which invoke the tunneling of two electrons through the NS interface.
Indeed, because our aim is to show that Andreev reflection can be used to measure noise,
we need to examine all two electron processes, we start with the transfer
of two electrons as quasiparticles above the gap.
Calculations of the matrix element in Eq. (\ref{rate}) are then carried out to second order in the
tunneling Hamiltonian using the T--matrix:
\begin{equation}
I=4\pi e\sum_f {|\langle f| H_T\frac{1}{i\eta-H_0+\epsilon_i}H_T |i \rangle|}^2 \delta(\epsilon_i-\epsilon_f)~,
\end{equation}
The initial state is a product:
\begin{equation}
|i\rangle=|G_L\rangle\otimes|G_S\rangle\otimes|R\rangle~,
\label{initial_state}
\end{equation}
where $|G_L\rangle$ denotes a ground state,  which corresponds to a filled
Fermi sea for the normal electrode.
$|G_S\rangle$ is the BCS ground state in superconductor lead.
$|R\rangle$ denotes the initial state of the environment.
On the other hand, our guess for the final state
should read
\begin{equation}
|f\rangle=
c^\dagger_{k,\sigma}c^\dagger_{k^\prime,\sigma^\prime}\gamma^\dagger_{q,\sigma}\gamma^\dagger_{q^\prime,\sigma^\prime}
|G_L\rangle\otimes|G_S\rangle\otimes|R^\prime\rangle~,
\label{guessNS1}\end{equation}
 when two electrons are emitted from superconductor. The ``guess'' of Eq. (\ref{guessNS1}) is an informed one: an electron pair is broken in the superconductor, and one electron tunnels to the normal metal lead, while the other becomes a quasiparticle in the superconductor; the same process for the second electron tunneled to the normal metal lead.
When superconductor lead absorbs two electrons
\begin{equation}
|f\rangle=c_{k,\sigma}c_{k^\prime,\sigma^\prime}\gamma^\dagger_{q,\sigma}\gamma^\dagger_{q^\prime,\sigma^\prime}
|G_L\rangle\otimes|G_S\rangle\otimes|R^\prime\rangle~,
\label{guessNS2}\end{equation}
The ``guess'' of Eq. and (\ref{guessNS2}) is: two electrons can tunnel from the normal metal lead and become quasiparticles in the superconductor.
Here, $|R^\prime\rangle$ is the final  state of the environment.

Introducing the closure relation for the eigenstates of the non-connected system $\{|\upsilon_i \rangle\}$, and using the fact that $\langle\upsilon|(\epsilon_i-H_0\pm i\eta)^{-1}|\upsilon\rangle= \mp i\int_{0}^{\infty}dt e^{i(\epsilon_i-\epsilon_\upsilon\pm i\eta)t}$, one can exponentiate all the energy denominators. Then, by transforming the time dependent phases into a time dependence of the tunneling Hamiltonian, we can integrate out all final and virtual states. This allows to rewrite the tunneling current in terms of tunneling operators in the interaction representation, to lowest order $O(T_0^4)$.

For the case of two electrons tunneling from Superconductor lead to normal metal lead,  the current reads
\begin{eqnarray}
I_\leftarrow&=&2e\int_{-\infty}^{\infty}\!\!\! dt\int_{0}^{\infty}\!\!\! dt^\prime\int_{0}^{\infty}\!\!\! dt^{\prime\prime} e^{-\eta(t^{\prime}+t^{\prime\prime})}e^{i(\mu_S-\mu_L)(2t-t^{\prime}-t^{\prime\prime})}\nonumber\\
&&\times\langle H_{T}^{\dagger}(t-t^{\prime\prime})H_{T}^{\dagger}(t)H_{T}(t^\prime)H_{T}(0)\rangle~.
\end{eqnarray}
Here we consider quasiparticle tunneling so
\begin{eqnarray}
&&\langle c^\dagger_{q_1\sigma_1}(t-t^{\prime\prime})c^\dagger_{q_2\sigma_2}(t)c_{q_3\sigma_3}(t^\prime)c_{q_4\sigma_4}(0)\rangle\nonumber\\
&=&-|v_q|^2|v_{q^\prime}|^2e^{-iE_{q}(t-t^{\prime\prime}-t^\prime)}e^{-iE_{q^\prime}t}\nonumber\\
&&+|v_q|^2|v_{q^\prime}|^2e^{-iE_{q}(t-t^{\prime\prime})}e^{-iE_{q^\prime}(t-t^\prime)}~.
\end{eqnarray}
The exponentials of phase operators are calculated with the definition of Eq. (\ref{jt}), so that \cite{recher_lossPRL,safi_devillard_martin}
\begin{eqnarray}
&&\!\!\!\!\!\!\!\!\!\!\!\langle e^{i\phi(t-t^{\prime\prime})}e^{i\phi(t)}e^{-i\phi(t^{\prime})}e^{-i\phi}\rangle\nonumber\\
\!\!\!\!\!\!\!\!&=&e^{J(t-t^{\prime\prime}-t^\prime)+J(t-t^{\prime\prime})+J(t-t^\prime)+J(t)-J(-t^{\prime\prime})-J(t^\prime)}\!~,
\end{eqnarray}
then we further assume that $J(t)\ll 1$, which means a low trans-impedance
approximation together with the fact that $J(t)$ is well behaved at large times. This allows to expand
the exponential of phase correlators $e^{J(t)}\approx 1+J(t)$.
The result for the current contains both an elastic and an inelastic contribution $I_\leftarrow=I_\leftarrow^{el}+I_\leftarrow^{inel}$, where
\begin{equation}
\!\!\!\!\!\!\!\!\!\!\!I_{\leftarrow}^{el}\!\simeq\!\frac{e\mathcal{T}^2}{16\pi^3R_K}\!\left\{\Psi_{0\leftarrow}-\frac{2\pi}{R_K}\!\int_{-\infty}^{\infty}\!\!\!\!\!\!\!\!\!d\omega\frac{|Z(\omega)|^2}{\omega^2}S_I(\omega)K_{2e\leftarrow}^{el}(\omega,eV,\eta)\right\}\!~,
\end{equation}
with $\Psi_{0\leftarrow}$, $K_{2e\leftarrow}^{el}(\omega,eV,\eta)$ are defined as in Eqs. (\ref{2epsi}) and (\ref{2eKel}) in Appendix A,
and
\begin{equation}
\!\!\!\!\!\!\!\!\!\!I_{\leftarrow}^{inel}\!\simeq\!\frac{e\mathcal{T}^2}{16\pi^2 R_K}\!\int^{\infty}_{2\Delta+2eV}\!\!\!\!\!\!\!\!\!\!\!d\Omega\frac{|Z(-\Omega)|^2}{\Omega^2}S_I(-(\Omega))K_{2e\leftarrow}^{inel}(\Omega,eV,\eta)\!~,
\end{equation}
where $K_{2e\leftarrow}^{inel}(\Omega,eV,\eta)$ is defined in Eq. (\ref{2eKinel}) in Appendix A.
The elastic contribution exists only if $eV<-\Delta$. For $-\Delta<eV$ only the inelastic part contributes to $I_\leftarrow$.

Similarly we calculate for $I_\rightarrow$. There is a symmetry between the magnitude between the right and left moving current
upon bias reversal: the expression for  $I_\leftarrow$ is the same as $I_\rightarrow$, if we replace $-eV$ by $eV$.

So in the interval $|eV|\le \Delta$, we obtain
\begin{eqnarray}
\Delta I_{PAT}(eV)\!\!\!&=&C_{2e}\int^{\infty}_{2\Delta-2eV}\!\!\!\!\!\!\!\!\!\!\!d\Omega\frac{|Z(-\Omega)|^2}{\Omega^2}S^+_{ex}(\Omega)K_{2e\rightarrow}^{inel}(\Omega,eV,\eta)\nonumber\\
&&\!\!\!\!\!\!\!\!-C_{2e}\int^{\infty}_{2\Delta+2eV}\!\!\!d\Omega\frac{|Z(-\Omega)|^2}{\Omega^2}S^+_{ex}(\Omega)K^{inel}_{2e\leftarrow}~.
\end{eqnarray}
with $K_{2e\rightarrow}^{inel}(eV)=K^{inel}_{2e\leftarrow}(-eV)$ and $C_{2e}=e\mathcal{T}^2/16\pi^2 R_K$.

\subsection{Two electron tunneling as a Cooper pair: Andreev reflection}
\label{ns_andreev_section}

In this case, we also needs to carry out calculations of the matrix element in Eq. (\ref{rate}) to second order in the
tunneling Hamiltonian. Typically, the initial state will be as shown in Eq. (\ref{initial_state}). 
On the other hand our guess for the final state reads:
\begin{equation}
|f\rangle=2^{-1/2}
[c^\dagger_{k,\sigma}c^\dagger_{k',-\sigma}-c^\dagger_{k',\sigma}c^\dagger_{k,-\sigma}]
|G_L\rangle\otimes|G_S\rangle\otimes|R^\prime\rangle~,
\label{guessNSA1}\end{equation}
 when a Cooper pair is emitted from superconductor, or
\begin{equation}
|f\rangle=2^{-1/2}
[c_{k,\sigma}c_{k',-\sigma}-c_{k',\sigma}c_{k,-\sigma}]
|G_L\rangle\otimes|G_S\rangle\otimes|R^\prime\rangle~,
\label{guessNSA2}
\end{equation}
when Superconductor lead absorbs a Cooper pair.
Here, $|R^\prime\rangle$ is the final  state of the environment.
The ``guess'' of Eqs. (\ref{guessNSA1}) and (\ref{guessNSA2}) is again an informed one:
indeed, the s-wave symmetry of the superconductor imposes that only singlet pairs of electrons
can be emitted or absorbed. This phenomenon has been described in the early work on entanglement in mesoscopic
physics \cite{recher_sukhorukov_loss,lesovik_martin_blatter}, and the resulting
final state can in principle be detected through a violation of Bell inequalities
\cite{chtchelkatchev}.
For this Andreev process,
we can now write the tunneling current as a function of the normal (and anomalous) Green's functions
of the normal metal lead, $G_{L\sigma}$, the quantum dot, $G_{D\sigma}$, and the superconductor, $F_{\sigma}$ (see Appendix B), which is the same as in Ref. \onlinecite{recher_lossPRL}
\begin{eqnarray}
I_\leftarrow&=&2eT_0^4\int_{-\infty}^{\infty}\!\!\! dt\int_{0}^{\infty}\!\!\! dt^\prime\int_{0}^{\infty}\!\!\! dt^{\prime\prime} e^{-\eta(t^{\prime}+t^{\prime\prime})}e^{i(\mu_S-\mu_L)(2t-t^{\prime}-t^{\prime\prime})}\nonumber\\
&&\!\!\!\!\!\!\!\!\!\!\!\!\!\!\!\!\!\!\times\!\!\!\!\!\!\sum_{k,k^\prime,q,q^\prime,\sigma}\!\!\!\!\!\!\left\{-G^>_{L\sigma}(k,t-t^{\prime\prime}-t^\prime)G^>_{L-\sigma}(k^\prime,t)F^*_{\sigma}(q^\prime,-t^{\prime\prime})F_{-\sigma}(q,t^{\prime})\right.\nonumber\\
&&\!\!\!\!\!\!\!\!\!+\left. G^>_{L\sigma}(k,t-t^{\prime\prime})G^>_{L-\sigma}(k^\prime,t-t^\prime)F^*_{\sigma}(q^\prime,-t^{\prime\prime})F_{\sigma}(q,t^{\prime})\right\}\nonumber\\
&&\!\!\!\!\!\!\!\!\times e^{J(t-t^{\prime\prime}-t^\prime)+J(t-t^{\prime\prime})+J(t-t^\prime)+J(t)-J(-t^{\prime\prime})-J(t^{\prime})}~.
\end{eqnarray}
The result for the current contains both an elastic and an inelastic contribution:
\begin{equation}
I_\leftarrow=I_{\leftarrow}^{el}+I_{\leftarrow}^{inel}~,
\end{equation}
where the elastic contribution reads
\begin{eqnarray}
I_{\leftarrow}^{el} &\simeq&\frac{e\mathcal{T}^2}{2\pi^3}\int_{eV}^{-eV}\!\!\! d\epsilon \int_{\Delta}^{\infty}\!\!\! dE\int_{\Delta}^{\infty}\!\!\! dE^\prime\frac{\Delta^2}{\sqrt{E^2-\Delta^2}\sqrt{{E^\prime}^2-\Delta^2}}\nonumber\\
&&\times\left\{ \left[1-\frac{4\pi}{R_K}\int_{-\infty}^{+\infty}d\omega\frac{|Z(\omega)|^2}{\omega^2}S_I(\omega)\right]\frac{1}{D^0_\leftarrow}\right.\nonumber\\
&&-\left.\frac{2\pi}{R_K}\int_{-\infty}^{+\infty}d\omega\frac{|Z(\omega)|^2}{\omega^2}\frac{S_I(\omega)}{D^{el}_\leftarrow}\right\}~,
\end{eqnarray}
The inelastic contribution to $I_{\leftarrow}$ is
\begin{eqnarray}
I_{\leftarrow}^{inel}&\simeq&\frac{e\mathcal{T}^2}{\pi^2 R_K}\int^{\infty}_{eV}\!\!\!d\epsilon\int^{\infty}_{eV}\!\!\!d\epsilon^\prime\int_{\Delta}^{\infty}\!\!\!dE\int_{\Delta}^{\infty}\!\!\!dE^\prime\nonumber\\
&&\!\!\!\!\!\!\!\!\!\!\!\!\!\!\!\!\!\!\!\!\!\!\!\!\!\!\!\!\!\!\times\frac{\Delta^2}{\sqrt{E^2-\Delta^2}\sqrt{{E^\prime}^2-\Delta^2}}\frac{|Z(-(\epsilon+\epsilon^\prime))|^2}{(\epsilon+\epsilon^\prime)^2}\frac{S_I(-(\epsilon+\epsilon^\prime))}{D^{inel}_\leftarrow}\!~.
\end{eqnarray}
where the denominators are specified in Appendix C.
$I_\rightarrow$ is derived in a similar manner, but its expression is omitted here. Nevertheless,
it effects will be displayed in the measurement of the noise of a point contact.

The above expressions constitute the central result of this work: one understands now how
the current fluctuations in the neighboring mesoscopic circuit give rise to inelastic and elastic
contributions in the current between a normal metal and a superconductor.

We find that both current contributions have the same form and the current fluctuations of the mesoscopic device
affect  the current in the detector at the energy corresponding to the total energy of two electrons in the normal lead.
Andreev reflection therefore acts as an energy filter.

Next, we can change variables as in the previous sections.
With an arbitrary bias $eV$, we obtain
\begin{eqnarray}
&&\!\!\!\!\!\!\!\!\!\!\!\!\!\Delta I_{PAT}(eV)\nonumber\\
&\!\!\!\!\!\!\!\!\!\!=&-C_{NS}\int_{-\infty}^{+\infty}d\omega\frac{|Z(\omega)|^2}{\omega^2}S^+_{excess}(-\omega)K^{el}_{NS}(\omega,eV,\eta)\nonumber\\
&&\!\!\!\!\!\!\!\!\!\!-\frac{C_{NS}}{2}\int^{2eV}_{-\infty}\!\!\! d\Omega\frac{|Z(\Omega)|^2}{\Omega^2}S^+_{excess}(-\Omega)K^{inel}_{NS}(\Omega,eV,\eta)\nonumber\\
&&\!\!\!\!\!\!\!\!\!\!-\frac{C_{NS}}{2}\int_{2eV}^{\infty}\!\!\! d\Omega \frac{|Z(-\Omega)|^2}{\Omega^2}S^+_{excess}(\Omega)K^{inel}_{NS}(\Omega,eV,\eta)~,
\label{PAT_Cooper}
\end{eqnarray}
with $C_{NS}=e\mathcal{T}^2\Delta^2/\pi^2 R_K$. The kernel functions $K^{el}_{NS}$, $K^{inel}_{NS}$ are shown in Appendix C.

\subsection{Quantum Point Contact as a source of noise}

\subsubsection{Spectral density of excess noise}

In this section, we illustrate the present results with a simple example. We consider for this purpose
a quantum point contact, which is a device whose noise spectral density is well characterized by using the scattering theory\cite{yang,lesovik}. Here however, we consider unsymmetrized noise correlators:
\begin{widetext}
\begin{eqnarray}
S^{+}(\omega)=\Bigg\{\begin{array}{cc}
                                   \frac{2e^2}{\pi}T(1-T)(eV_d-\hbar\omega)\theta(eV_d-\hbar\omega)~, & \textrm{if $\omega\ge 0$},\\
                                   \frac{2e^2}{\pi}[-2T^2\hbar\omega - T(1-T)(eV_d+\hbar\omega)\theta(-eV_d-\hbar\omega) + T(1-T)(eV_d-\hbar\omega)]~, & \textrm{if $\omega < 0$}~,
                              \end{array}
\end{eqnarray}
\end{widetext}
where $\theta$ is the Heaviside function and $T$ is the transmission probability, which is assumed to have a weak dependence on
energy over the voltage range $eV$.

As pointed out above, we are computing the difference of the PAT currents in the presence and in the absence of the DC
bias. This means that we insert the spectral density of excess noise of the mesoscopic device, which for a point contact
bears most of its weight near zero frequencies.
Excess noise decreases linearly to zero over a range $[0,\pm eV_d]$ for positive and negative frequencies:
\begin{eqnarray}
S^{+}_{ex}(\omega)\!=\!
\frac{2e^2}{\pi}T(1-T)(eV_d-|\hbar\omega|)\theta(eV_d-|\hbar\omega|)\!~,
\label{excess_noise}
\end{eqnarray}

\subsubsection{Numerical calculations}

\begin{figure*}[t]
\centerline{\includegraphics[width=12.cm]{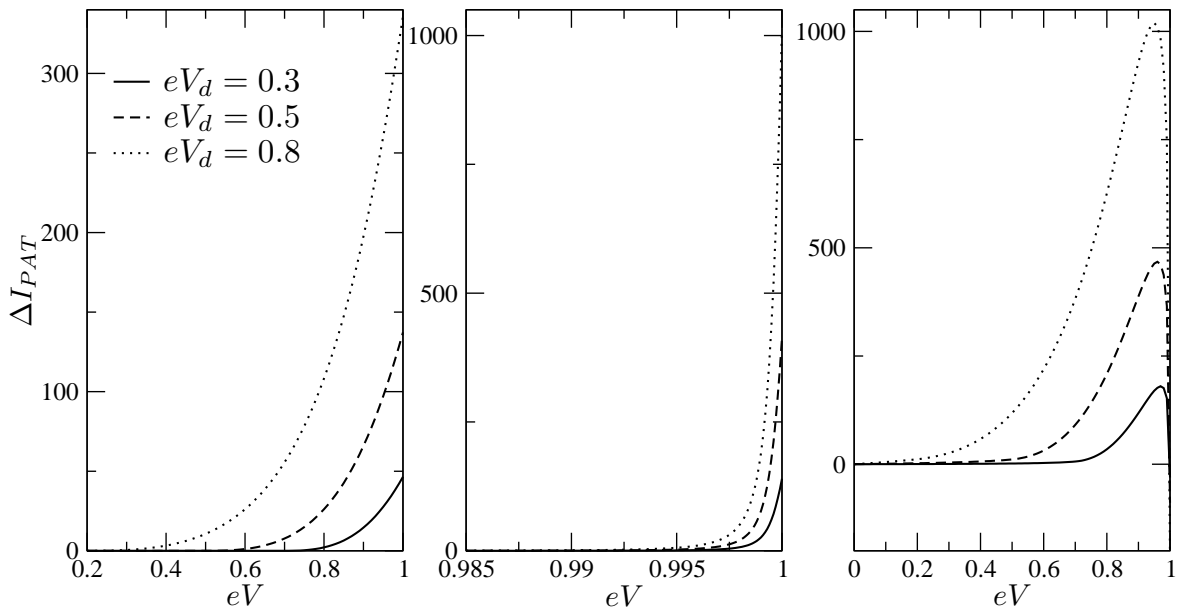}}
\caption{ $\Delta I_{PAT}$ plotted as a function of the DC bias voltage, for a mesoscopic device voltage bias $eV_d$: $0.3$ (continuous line), $0.5$ (dashed line) and $0.8$ (dotted line). The  left/center/right panels depict single quasiparticle tunneling/two quasiparticle tunneling/Andreev reflection.
$\Delta I_{PAT}$ is in units of $C_{2e}$, with $\mathcal{T}=0.6$ (see text).}
\label{figNSt1}
\end{figure*}
We choose a generic form for the transimpedance, similar in spirit to that chosen in Ref. \onlinecite{aguado_kouwenhoven}.
Considering the circuit in Figs. \ref{figNS} and \ref{figNDS}, at $\omega=0$, device and detector are not coupled and the transimpedence should therefore vanish.
On the other hand, the transimpedance is predicted to have a constant behavior at large frequency.
We therefore choose the following generic form for the transimpedance
\begin{equation}
|Z(\omega)|^2=\frac{(R\omega)^2}{\omega_{0}^{2}+\omega^2}~,
\end{equation}
where $R$ is the typical high frequency impedance and the crossover frequency $\omega_{0}$ is estimated
from the experimental data of Ref. \onlinecite{deblock2}, choosing a finite cutoff frequency $\omega_0$ means that
at frequencies $\omega\ll\omega_0$, the mesoscopic circuit has no influence on the
detector circuit because low frequencies do not propagate through a capacitor.

We calculate numerically the PAT currents in the three above cases: single and two quasiparticle tunneling,
and Andreev reflection. All energies are measured with respect to the superconducting gap $\Delta$.
In these units we chose $\omega_0=0.3$ and $\eta=0.001$.
Currents are typically plotted as a function of the DC bias voltage $eV$ of the detector, for several
values of the mesoscopic bias voltage $eV_d$ (the ``environment'').
Our motivation is to consider the PAT currents with
the condition $|eV|<1$ ($|eV|<\Delta$), where the effect of the environment on the PAT current is most pronounced,
and we shall predict that two electrons tunneling as a Cooper pair (Andreev processes) contributes the most
to the PAT current, except close to $eV=\Delta$. Because of the symmetry between negative $eV$ and positive
$eV$, we display the results for $eV>0$.
The PAT currents for the three above processes are plotted next to one other in Fig. \ref{figNSt1} for comparison.

We find in Fig. \ref{figNSt1} that if the chemical potential of the
normal metal lead is close to the potential of the superconductor
lead ($eV\ll\Delta$), the PAT currents in the three cases are
suppressed. For the two cases of quasiparticle tunneling, the PAT
currents are equal to zero below a certain threshold. The single
quasiparticle current differs from zero at a threshold which is
identified as $\Delta -eV_d$, that is quasiparticles are able to
tunnel above that superconductor gap only if they can borrow the
necessary energy from the mesoscopic device. This explains why the
curves associated with different values of the mesoscopic device
bias voltage are shifted to the right as $eV_d$ decreases. For two
quasiparticle particle tunneling, we observe that the PAT current
has a similar threshold, which, compared to Fig. \ref{figNSt1}a, is
pushed toward the right in Fig. \ref{figNSt1}b, because more energy
is needed to transfer two electrons above the gap, compared to a
single electron. Nor surprisingly, the curves corresponding are once
again shifted to the right with decreasing $eV_d$. These curves all
have a sharp maximum at $eV=\Delta$.

We turn now to the Andreev PAT difference current, which dominates the two above processes at small
and moderate bias.
Note that the total Andreev current contains an elastic contribution as well as an inelastic contribution below the gap,
contrary to quasiparticle tunneling which have contributions below the gap only because these processes are
photo-assisted.
Because we are computing the difference between PAT current with and without the mesoscopic bias voltage,
we expect that the elastic contribution cancels out. However the first term of Eq. (\ref{PAT_Cooper})
tells us that the presence of the environment also gives rise to an effective elastic contribution
to the PAT Andreev difference current.
Unfortunately, this elastic correction is not small compared to the true Andreev current.

Looking at Fig. \ref{figNSt1},
we note that the PAT curve for Andreev processes is shifted
to the left when the bias of the mesoscopic device is increased.
The environment provides/absorbs energy to pairs of electrons whose energies
are not symmetrical with respect to the superconducting chemical potential.
At very weak $eV$, the elastic correction
is small compared to the true Andreev current.  We expect the PAT current contributions to originate from
pairs of electrons in the normal metal below the superconducting chemical potential which can extract
a photon from the environment. At the same time, Cooper pairs can be ejected in the normal metal
as a reverse Andreev processes provided they borrow a photon to the environment.
There is a balance between the right and left currents at very weak bias.

As the bias is increased, electrons pairs whose energies are
above this chemical potential will now be able to yield a photon to the environment,
giving rise to an increase of the inelastic PAT current.
Also, the reverse Andreev process mentioned above becomes more restricted
because the available empty states for electrons in the leads lie higher at
positive bias. The elastic PAT current (not shown in the figures) increases when we increase the detector bias but it is always smaller than the inelastic contribution. The total PAT current increases gradually (Fig. \ref{figNSt1}, right panel).

\begin{figure}[t]
\centerline{\includegraphics[width=7.cm]{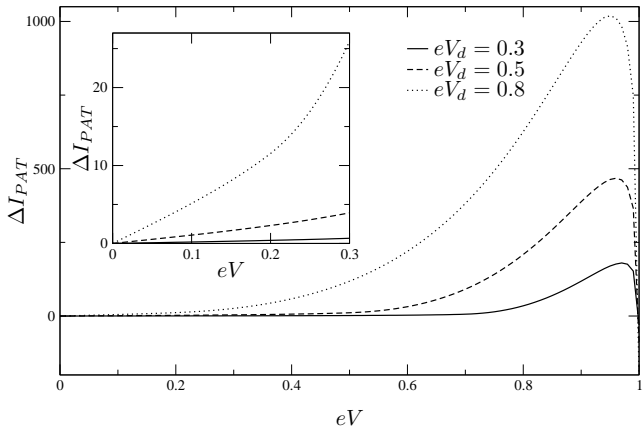}}
\caption{Andreev reflection contribution to $\Delta I_{PAT}$ plotted as a function of the DC bias voltage, for a mesoscopic device voltage bias $eV_d$: $0.3$ (continuous line), $0.5$ (dashed line) and $0.8$ (dotted line).
Same units as in Fig. \ref{figNSt1}. The inside panel is a zoom of the same contribution at small $eV$,
displaying linear behavior.}
\label{fig_Cooper}
\end{figure}

\begin{figure}[t]
\centerline{\includegraphics[width=7.cm]{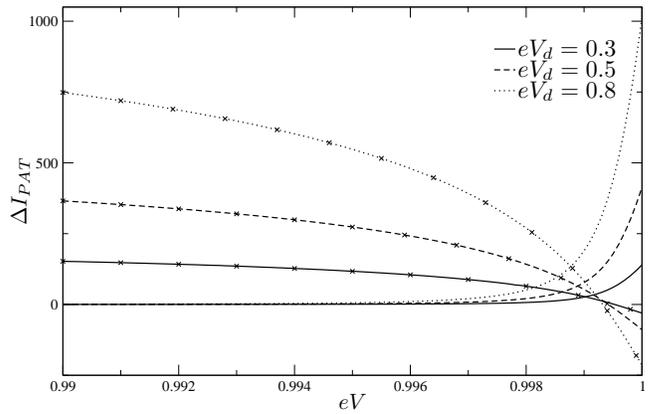}}
\caption{Crossover between Andreev reflection (crossed line) and two quasiparticle tunneling current (uncrossed line)
close to the gap, for a mesoscopic device voltage bias $eV_d$: $0.3$ (continuous line), $0.5$ (dashed line) and $0.8$ (dotted line).
Same units as in Fig. \ref{figNSt1}.}
\label{crossover2e}
\end{figure}

A zoom of this Andreev contribution is made in the region of small bias. There is in fact no threshold
for the PAT Andreev current: for small bias, it
has a linear behavior (Fig. \ref{fig_Cooper}).
According to Eq. (\ref{PAT_Cooper}), when $eV\ge eV_d/2$, the inelastic difference current of a Cooper pair tunneling from S to L by absorbing energy from mesoscopic device vanishes. The PAT current now describes two electron tunneling from L to S elastically or inelastically. In this intermediate bias regime, the elastic and inelastic contributions have now a tendency to cancel each other. The current reaches a maximum close to the gap, and then it decreases dramatically at the gap.
This is consistent with the fact that for positive bias, the initial state for two
quasiparticle tunnel processes and for Andreev reflection is precisely the same:
close to the gap two quasiparticle tunneling takes over the Andreev process.
It becomes more efficient for electrons to be activated above the gap than to
be converted into a Cooper pair because the energy loss needed for the latter
is quite large.
Unlike a conventional normal metal -- superconductor junction with elastic scattering only, where the relative
importance of quasiparticle tunneling and Andreev
reflection are interchanged precisely at the gap,
here the dominance of quasiparticle tunneling manifests itself
before the voltage bias reaches the gap. Note also in Fig. \ref{figNSt1}, that the magnitude of the Andreev current
before the two quasiparticle threshold is precisely the same as the magnitude of the two quasiparticle at
$eV=\Delta$, which confirms this conversion scenario. A comparison of the two processes is displayed
in Fig. \ref{crossover2e}.

In practice, the different contributions to the PAT current cannot be separated:
one measures the sum of the three contributions which are plotted in Fig. \ref{figNSt1}.
However, we claim that for a broad voltage range (from $eV=0$ to the two quasiparticle threshold),
the main contribution to the current comes from photo-assisted Andreev processes.
The confrontation of Eq. (\ref{PAT_Cooper}) with an experimental measurement of the PAT current
below the gap could thus serve an effective noise measurement, as the weight functions
$K^{el}_{NS}(\omega,eV,\eta)$ and $K^{inel}_{NS}(\Omega,eV,\eta)$
are known.

Notice that in all our numerical results, the PAT currents are plotted in units of
$e^3R^2\mathcal{T}^2 T(1-T)\Delta/8\pi^3\hbar^2 R_K$.
We put some tentative numbers in these quantities.
Here $\mathcal{T}=0.6$ is the
effective transmission coefficient of the NS interface, $\Delta=240\mu eV$, $T=0.5$ is the transmission of the
quantum point contact to be measured, $R=0.03R_K$ ($R_K$ is the resistance quantum) is the resistance
which enters the transimpedance. This implies, e.g. for the PAT current in Fig. \ref{fig_Cooper}
at the top of the peak, $\Delta I_{PAT}\simeq 10^{-10} A$ with the device bias $V_{d}=0.8\Delta/e=48\mu V$, which seems
an acceptable value which is compatible with current measurement techniques.

\section{Tunneling current through a NDS junction}

We now turn to a different setup for noise detection where electrons in a normal metal lead transit through a quantum dot in the Coulomb blockade regime before going into the superconductor. The essential ingredients are the same as the previous
section, except that additional energy filtering occurs because of the dot. In this section, we choose the parameters
of the device so that only Andreev processes are relevant.

\subsection{Model Hamiltonian}

The Hamiltonian which describes the decoupled normal metal lead -- dot -- superconductor -- environment
(mesoscopic circuit) system reads
\begin{equation}
H_0=H_{0_L}+H_{0_D}+H_{0_S}+H_{env}~,
\end{equation}
where the Hamiltonian of normal metal lead and superconductor lead are described as above.

The Hamiltonian for the quantum dot reads
\begin{eqnarray}
H_{0_D}&=&\sum_{\sigma}\epsilon_D c^\dagger_{D,\sigma}c_{D,\sigma}+Un_\uparrow n_\downarrow~,
\end{eqnarray}
where $U$ will be assumed to be infinite, assuming a small capacitance of the dot. We consider that the dot possesses
only a single energy level for simplicity.

The tunneling Hamiltonian includes electron tunneling between the superconductor
and the dot, as well as the tunneling between the dot and the normal metal lead
\begin{eqnarray}
H_T&=&(H_{T1}+H_{T2})+h.c~,\\
H_{T1}&=&\sum_{q,\sigma}T_{D,q}c^\dagger_{D,\sigma}c_{q,\sigma}e^{-i\phi}\nonumber ~,\\
H_{T2}&=&\sum_{k,\sigma}T_{k,D}c^\dagger_{k,\sigma}c_{D,\sigma}~,\nonumber
\end{eqnarray}
the indices k, D, q refer to the normal metal lead, quantum dot, superconductor.
 We consider the simple case $T_{D,q}=T_1$, and $ T_{k,D}=T_2$.

For photo-assisted Andreev processes,
one needs to carry out calculations of the matrix element in Eq. (\ref{rate}) to fourth order in the
tunneling Hamiltonian.
In what follows, we assume that the dot is initially empty, owing to the
asymmetry between the two tunnel barriers. The barrier between the normal metal lead
and the dot is supposed to be opaque compared to that between the dot and the
superconductor. As a result, the rate of escape of
electrons from the dot to the superconductor is substantially larger than the tunneling
rate of electrons from the normal lead to the dot (see below for actual numbers).

There are two possibilities for charge transfer processes:
a Cooper pair in the superconductor is transmitted to the normal lead or vice versa.
The first process involves the electron from a Cooper pair tunneling
onto the dot, next this electron escapes in the lead; the other electron from the same Cooper pair then
undergoes the same two tunneling processes. Similar transitions, in the opposite direction,
are necessary for two electrons from the normal lead to end up as a Cooper pair in the superconductor.
Note that this description of events assumes implicity that superconductor lead remains in the ground state
in the initial and final states (Andreev process). On the other hand, if normal metal lead is initially in the ground
state (filled Fermi sea),
it is left in an excited state with two electrons having energies above Fermi energy $E_F$
in the final state. The extra energy has been provided by the environment.
Typically,
\begin{equation}
|i\rangle=|G_L\rangle\otimes|G_S\rangle\otimes|0_{QD}\rangle\otimes|R\rangle~,
\end{equation}
where $|G_{\ldots}\rangle$ denotes a ground state,  which corresponds to a filled
Fermi sea for the normal electrode. $|0_{QD}\rangle$ is the vacuum of the quantum dot, and
$|R\rangle$ denotes the initial state of the environment. On the other hand our guess for the final
state
should read
\begin{equation}
|f\rangle=2^{-1/2}
[c^\dagger_{k,\sigma}c^\dagger_{k',-\sigma}-c^\dagger_{k',\sigma}c^\dagger_{k,-\sigma}]
|G_L\rangle\otimes|G_S\rangle\otimes|0_{QD}\rangle\otimes|R^\prime\rangle~,
\label{guess1}\end{equation}
 when a Cooper pair is emitted from S, or
\begin{equation}
|f\rangle=2^{-1/2}
[c_{k,\sigma}c_{k',-\sigma}-c_{k',\sigma}c_{k,-\sigma}]
|G_L\rangle\otimes|G_S\rangle\otimes|0_{QD}\rangle\otimes|R^\prime\rangle~,
\label{guess2}\end{equation}
when superconductor lead absorbs a Cooper pair.
Here, $|R^\prime\rangle$ is the final  state of the environment.
The justification for the choice of Eqs. (\ref{guess1}) and (\ref{guess2})
is the same as in Sec. \ref{ns_andreev_section}.

\subsection{General formula for the photo-assisted Andreev current}

For the case of two electrons tunneling from superconductor lead to normal metal lead,  the current reads
\begin{widetext}
\begin{eqnarray}
I_\leftarrow&=&2e\int_{-\infty}^{\infty}\!\!\! dt\int_{0}^{\infty}\!\!\! dt_1\int_{0}^{\infty}\!\!\! dt_2\int_{0}^{\infty}\!\!\! dt_3\int_{0}^{\infty}\!\!\! dt_{1}^{\prime}\int_{0}^{\infty}\!\!\! dt_{2}^{\prime}\int_{0}^{\infty}\!\!\! dt_{3}^{\prime} e^{-\eta(t_1+t_2+t_3+t_{1}^{\prime}+t_{2}^{\prime}+t_{3}^{\prime})}\nonumber\\
&&\times e^{i\mu_S(2t-t^{\prime}_{1}-t^{\prime}_{2}-2t^{\prime}_{3}-t_1-t_2)}e^{-i\mu_L(2t-t^{\prime}_{2}-t^{\prime}_{3}-2t_1-t_2-t_3)}\nonumber\\
&&\!\!\!\!\!\!\!\!\!\!\!\!\times\langle H_{T1}^{\dagger}(t-t^{\prime}_{1}-t^{\prime}_{2}-t^{\prime}_{3}) H_{T2}^{\dagger}(t-t^{\prime}_{2}-t^{\prime}_{3}) H_{T1}^{\dagger}(t-t^{\prime}_{3})H_{T2}^{\dagger}(t) H_{T2}(t_1+t_2+t_3) H_{T1}(t_1+t_2)H_{T2}(t_1)H_{T1}(0)\rangle~.
\end{eqnarray}
\end{widetext}

The problem is thus reduced to the calculation of correlators of the tunneling Hamiltonian in the ground state.
Using Wick's theorem, these can be expressed in terms of single particle Green's function because the Hamiltonian
of the isolated components is quadratic (except, maybe for the environment, which is dealt separately).
We can now write the tunneling current as a function of the normal (and anomalous) Green's functions
of the normal metal lead, $G_{L\sigma}$, the quantum dot, $G_{D\sigma}$, and the superconductor, $F_{\sigma}$ (which are shown in Appendix B)
\begin{widetext}
\begin{eqnarray}
I_\leftarrow &=&2eT^{4}_{1}T^{4}_{2}\int_{-\infty}^{\infty}\!\!\! dt\int_{0}^{\infty}\!\!\! dt_1\int_{0}^{\infty}\!\!\! dt_2\int_{0}^{\infty}\!\!\! dt_3\int_{0}^{\infty}\!\!\! dt_{1}^{\prime}\int_{0}^{\infty}\!\!\! dt_{2}^{\prime}\int_{0}^{\infty}\!\!\! dt_{3}^{\prime} e^{-\eta(t_1+t_2+t_3+t_{1}^{\prime}+t_{2}^{\prime}+t_{3}^{\prime})}\nonumber\\
&&\times e^{i\mu_S(2t-t^{\prime}_{1}-t^{\prime}_{2}-2t^{\prime}_{3}-t_1-t_2)}e^{-i\mu_L(2t-t^{\prime}_{2}-t^{\prime}_{3}-2t_1-t_2-t_3)}\nonumber\\
&&\times\sum_{k,k^\prime,q,q^\prime,\sigma}\left[-F^{*}_{\sigma}(q^\prime,-t^{\prime}_{1}-t^{\prime}_{2})F_{-\sigma}(q,t_1+t_2)G^{>}_{L\sigma}(k,t-t^{\prime}_{2}-t^{\prime}_{3}-t_1-t_2-t_3)G^{>}_{L-\sigma}(k^\prime,t-t_1)\right.\nonumber\\
&&\times\left. G^{\tilde{t}}_{D\sigma}(-t^{\prime}_{1})G^{\tilde{t}}_{D-\sigma}(-t^{\prime}_{3})G^{t}_{D\sigma}(t_3)G^{t}_{D-\sigma}(t_1)+F^{*}_{\sigma}(q^\prime,-t^{\prime}_{1}-t^{\prime}_{2})F_{\sigma}(q,t_1+t_2)\right.\nonumber\\
&&\times\left. G^{>}_{L\sigma}(k,t-t^{\prime}_{2}-t^{\prime}_{3}-t_1)G^{>}_{L-\sigma}(k^\prime,t-t_1-t_2-t_3)G^{\tilde{t}}_{D\sigma}(-t^{\prime}_{1})G^{\tilde{t}}_{D-\sigma}(-t^{\prime}_{3})G^{t}_{D-\sigma}(t_3)G^{t}_{D\sigma}(t_1)\right]\nonumber\\
&&\times e^{J(t-t^{\prime}_{1}-t^{\prime}_{2}-t^{\prime}_{3})+J(t-t^{\prime}_{3})+J(t-t^{\prime}_{1}-t^{\prime}_{2}-t^{\prime}_{3}-t_1-t_2)+J(t-t^{\prime}_{3}-t_1-t_2)-J(-t^{\prime}_{1}-t^{\prime}_{2})-J(t_1+t_2)}~.
\end{eqnarray}
\end{widetext}
The result for the current contains both an elastic and an inelastic contribution. The elastic contribution reads
\begin{eqnarray}
I_{\leftarrow}^{el}\!&\simeq&\!\frac{e\gamma_{1}^{2}\gamma_{2}^{2}}{2\pi^3}\!\int_{eV}^{-eV}\!\!\!\!\!\!d\epsilon \int_{\Delta}^{\infty}\!\!\!\!dE\int_{\Delta}^{\infty}\!\!\!\! dE^\prime\frac{\Delta^2}{\sqrt{E^2-\Delta^2}\sqrt{{E^\prime}^2-\Delta^2}}\nonumber\\
&&\times\left\{ \left[1-\frac{4\pi}{R_K}\int_{-\infty}^{+\infty}d\omega\frac{|Z(\omega)|^2}{(\omega)^2}S_I(\omega)\right]\frac{1}{D^0_\leftarrow}\right.\nonumber\\
&&-\left.\frac{2\pi}{R_K}\int_{-\infty}^{+\infty}d\omega\frac{|Z(\omega)|^2}{(\omega)^2}\frac{S_I(\omega)}{D^{el}_\leftarrow}\right\}~,
\end{eqnarray}
with $D^0_\leftarrow$ is the original denominator which is not affected by the environment,
which is defined by Eq. (\ref{denominatorNDS1}),
and $D^{el}_\leftarrow$ is the denominator product affected by the environment (see Eq. (\ref{denominatorNDS2}) of Appendix D).
The inelastic contribution reads
\begin{eqnarray}
I_\leftarrow^{inel}&\simeq&\frac{e\gamma_1^2\gamma_2^2}{\pi^2 R_K}\int_{eV}^{\infty}\!\!\! d\epsilon\int_{eV}^{\infty}\!\!\! d\epsilon^\prime\int_{\Delta}^{\infty}\!\!\! dE\int_{\Delta}^{\infty}\!\!\! dE^\prime\,\,\,\,\,\,\nonumber\\
&&\!\!\!\!\!\!\!\!\!\!\!\!\!\!\!\!\!\!\!\!\!\!\!\!\!\!\!\!\!\!\!\!\times\frac{\Delta^2}{\sqrt{E^2-\Delta^2}\sqrt{{E^\prime}^2-\Delta^2}}\frac{|Z(-(\epsilon+\epsilon^\prime))|^2}{(\epsilon+\epsilon^\prime)^2}\frac{S_I(-(\epsilon+\epsilon^\prime))}{D^{inel}_\leftarrow}\!\!~,
\end{eqnarray}
where $D^{inel}_\leftarrow$ is the denominator product attributed to the inelastic current,
which is defined in
Eq. (\ref{denominatorNDS3}) of Appendix D.

Here, $\gamma_1=2\pi\mathcal{N}_{S}T_1^2$,
$\gamma_2=2\pi\mathcal{N}_{N}T_2^2$ define the tunneling rates
between the superconductor and the dot, between the dot and the
normal metal lead, respectively, with $\mathcal{N}_{S}$ and
$\mathcal{N}_{N}$ the density of states per spin of the two metals
in the normal state, at the chemical potentials $\mu_S$ and $\mu_L$,
respectively. All contributions to the current contain denominators
where the infinitesimal $\eta$ (adiabatic switching parameter) is
included in order to avoid divergences. In fact, it has been shown
in Refs. \onlinecite{recher_sukhorukov_loss,lehur} that
a proper resummation of the perturbation series, including all round
trips from the dot to the normal leads, leads to a broadening of the
dot level. We take into account this broadening by replacing $\eta$
by $\gamma/2$ with $\gamma=\gamma_1+\gamma_2$ into our calculations
(including only a broadening due to the superconductor). As
mentioned above, we have assumed that $\gamma_1\gg\gamma_2$. In
order to avoid the excitation of quasiparticles above the gap, these
rates need also to fulfill the condition $\epsilon_D+\gamma<\Delta$.
In what follows, we keep the notation $\eta$ in our expressions,
bearing in mind that it represents the line width associated with
the leads. For numerical purposes, it will be sufficient to assume
that $\eta$ is kept very small compared to the superconducting gap,
as well as all the relevant level energies (dot level position, bias
voltages, etc...).

The above expressions constitute the second main result of this work: one understands now how
the current fluctuations in the neighboring mesoscopic circuit give rise to inelastic and elastic
contributions in the current in the NDS device.

We find that both right and left current contributions have the same
form. The current - current fluctuations of the mesoscopic device
affect the detector current at the energy corresponding to the total
energy of two electrons exiting in (or entering from) the normal
lead. Therefore, we proceed to the same change of variables a for
the single NS junction.

For $eV>0$, elastic current contributions in $I_\rightarrow$ are present but the same contributions
in $I_\leftarrow$ vanish (the opposite is true for the case of $eV<0$).

Changing variables in the inelastic contributions, and defining the kernel functions $K^{el}_{NDS}(\omega,eV,\epsilon_D,\eta)$, $K^{inel}_{NDS}(\Omega,eV,\epsilon_D,\eta)$ as in Appendix D, for $eV>0$ and $eV<0$, we obtain:
\begin{eqnarray}
&&\!\!\!\!\!\!\!\!\!\!\!\!\!\!\!\Delta I_{PAT}(eV)\nonumber\\
\!\!&=&\!-C_{NDS}\int_{-\infty}^{+\infty}\!\!\!\!\!\!\!\!\!d\omega\frac{|Z(\omega)|^2}{\omega^2}S^+_{ex}(-\omega)K^{el}_{NDS}(\omega,eV,\epsilon_D,\eta)\nonumber\\
&&\!\!\!\!\!\!\!\!\!\!\!\!\!\!\!-\frac{C_{NDS}}{2}\int^{2eV}_{-\infty}\!\!\!\!\!\! d\Omega\frac{|Z(\Omega)|^2}{\Omega^2}S^+_{ex}(-\Omega) K^{inel}_{NDS}(\Omega,eV,\epsilon_D,\eta)\nonumber\\
&&\!\!\!\!\!\!\!\!\!\!\!\!\!\!\!-\frac{C_{NDS}}{2}\int_{2eV}^{\infty}\!\!\!\!\!\!d\Omega \frac{|Z(-\Omega)|^2}{\Omega^2}S^+_{ex}(\Omega) K^{inel}_{NDS}(\Omega,eV,\epsilon_D,\eta)~,
\label{I_PAT_NDS}
\end{eqnarray}
with $C_{NDS}=e\gamma_1^2\gamma_2^2\Delta^2/\pi^2 R_K$.

\begin{figure*}[t]
\centerline{\includegraphics[width=11.cm]{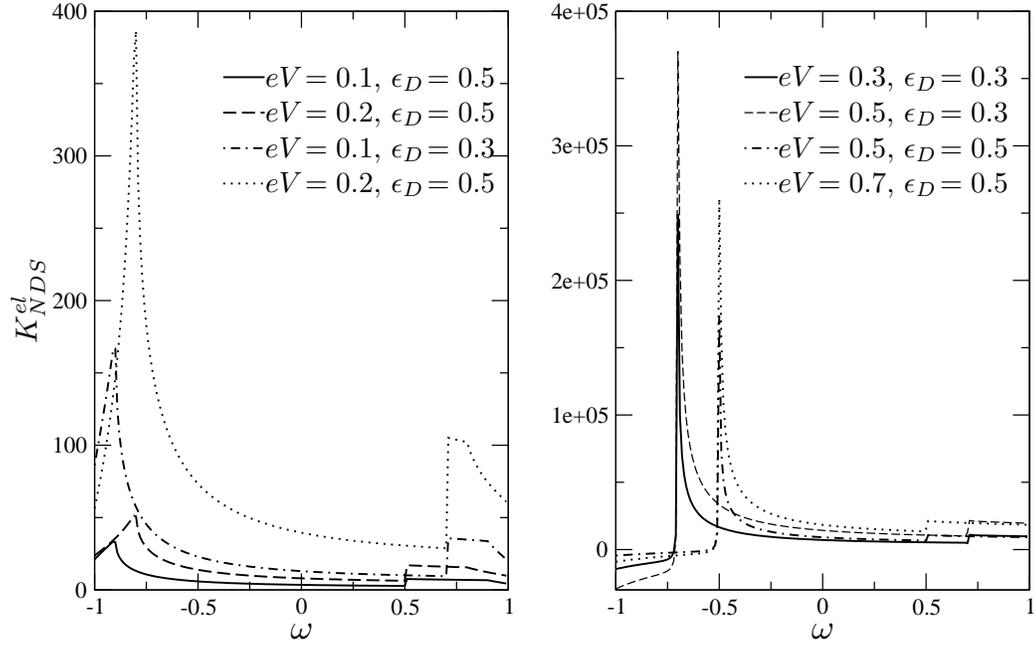}}
\caption{The weight function $K^{el}_{NDS}$ as a function of frequency $\omega$.
Right panel, $|eV|\ge\epsilon_D$; left panel $|eV|<\epsilon_D$.}
\label{KelNDS}
\end{figure*}
The first term in Eq. (\ref{I_PAT_NDS}) describes the elastic contribution in the PAT current. Although we are less interested in this contribution,  we cannot ignore it in practice because it contributes to the total $\Delta I_{PAT}$. The environment affects this current contribution but at the end of the tunneling processes, there is no energy exchanged between the device and the detector circuit.
The second term in Eq. (\ref{I_PAT_NDS}) describes the tunneling of a Cooper pair from the normal lead to the superconductor via the quantum dot, with energy exchange. The electrons can absorb energy (in the case their total energy is smaller than the superconductor chemical potential $\mu_S$) or emit energy (if their total energy is bigger than $\mu_S$).
The last term in Eq. (\ref{I_PAT_NDS}) describes the inverse tunneling process: a Cooper pair absorb energy
from the neighboring device, its constituent electrons then tunnel to the normal lead.
In this event, the total energy of the outgoing electrons is positive.
If on the contrary, this total energy is negative, the Cooper pair has
emitted energy to the device.

\begin{figure*}[t]
\centerline{\includegraphics[width=12.cm]{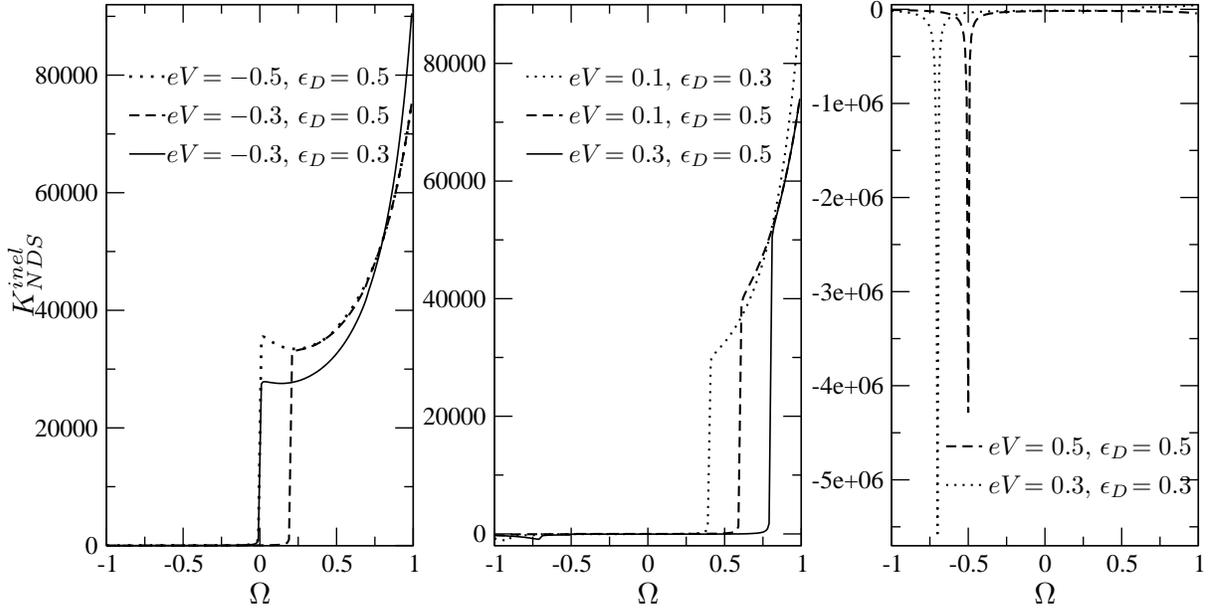}}
\caption{ the weight function $K^{inel}_{NDS}$ as a function of frequency $\Omega$.
Left panel, $eV<0$;  center panel $0<eV<\epsilon_D$; right panel $eV>\epsilon_D$}
\label{KinelNDS}
\end{figure*}

In order to understand how the detector circuit affects the behavior of the current (in the presence of the environment), we investigate the weight functions $K^{el}_{NDS}(\omega,eV,\epsilon_D,\eta)$ and $K^{inel}_{NDS}(\Omega,eV,\epsilon_D,\eta)$
separately.

The weight function $K^{el}_{NDS}(\omega,eV,\epsilon_D,\eta)$  is plotted in Fig. \ref{KelNDS} as a function of frequency $\omega$, for two values of the bias voltage and two values of the dot level position.
This elastic kernel is symmetric under bias voltage reversal ($K^{el}_{NDS}(-eV)=-K^{el}_{NDS}(eV)$). From the right panel of this figure, where we consider $|eV|\ge\epsilon_D$, we find that there is a small step at $\omega=\Delta-\epsilon_D$ and a sharp peak at $\omega=-\Delta+\epsilon_D$. The peak is assymetric, and its
height is much larger than that of the step. When $\omega<-\Delta+\epsilon_D$, $K^{el}_{NDS}$ changes sign and becomes negative. The voltage bias $eV$ mainly affects the amplitude of the peak and of the step in $K^{el}_{NDS}$.
The left panel
of Fig. \ref{KelNDS} describes $K^{el}_{NDS}$ when $|eV|<\epsilon_D$. The peak height decreases quite fast as a function of $eV$,
and its location is shifted at $\omega=-\Delta+eV$. The peak is symmetric for large bias.

We turn now to the truly photo-assisted processes, which involve either absorption or emission of energy.
The kernel $K^{inel}_{NDS}(\Omega,eV,\epsilon_D,\eta)$ is plotted in Fig. \ref{KinelNDS} as a function of frequency $\Omega$, which corresponds to the total energy of two electrons, for
$\epsilon_D>0$. In the left panel, $eV$ is negative, and in the center panel, $eV$ is positive but $eV<\epsilon_D$, and finally the right panel of Fig. \ref{KinelNDS} describes $eV\ge\epsilon_D$.
We find that when $eV<\epsilon_D$, there is a step at $\Omega=\epsilon_D+eV$. When we increase $eV$ close to
$\epsilon_D$, the step still dominates $K^{inel}_{NDS}$ but there is a small peak at $\Omega=-\Delta+eV$. When  $eV\ge\epsilon_D$ this (inverted) peak is very sharp. This is explicit in the right panel. The inverted peak, which has a large amplitude, makes it now difficult to observe the step. The (inverted) peak is located at $\Omega=-\Delta+\epsilon_D$. Again, $eV$ mostly affects the amplitude of $K^{inel}_{NDS}$. For $\epsilon_D<0$  (not shown), the result is similar to that of $\epsilon_D>0$ with opposite $eV$, but the amplitude of the peak is doubled compared to that of $\epsilon_D>0$ when $|eV|\ge|\epsilon_D|$.

Note that understanding the behavior of the two weight functions as a function of the different parameters
($eV$ and $\epsilon_D$) of the detector is crucial. It allows to control the effect of the device voltage bias $eV_d$ on the DC current of the detector, and it is therefore the key for extracting the noise from the measurement of this DC current.

\subsection{Application to a Quantum Point Contact}
\begin{figure*}[t]
\centerline{\includegraphics[width=11.cm]{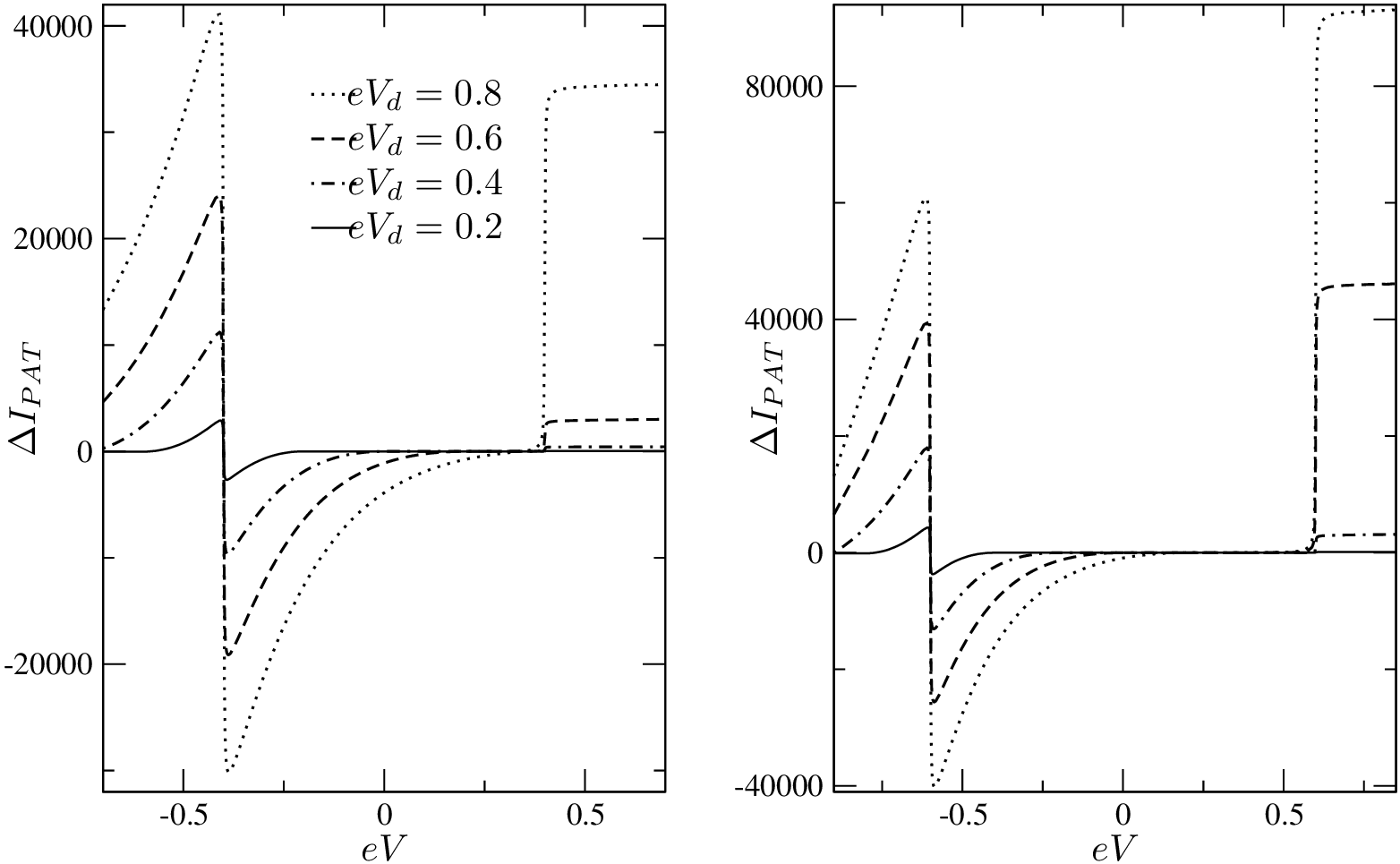}}
\caption{ $\Delta I_{PAT}$ plotted as a function of the detector bias voltage, for dot energy level: $\epsilon_d=0.4$
(left panel), $0.6$ (right panel) and for several values of device bias $eV_d$.}

\label{figIPAT_eV}
\end{figure*}

\begin{figure*}[t]
\centerline{\includegraphics[width=12cm]{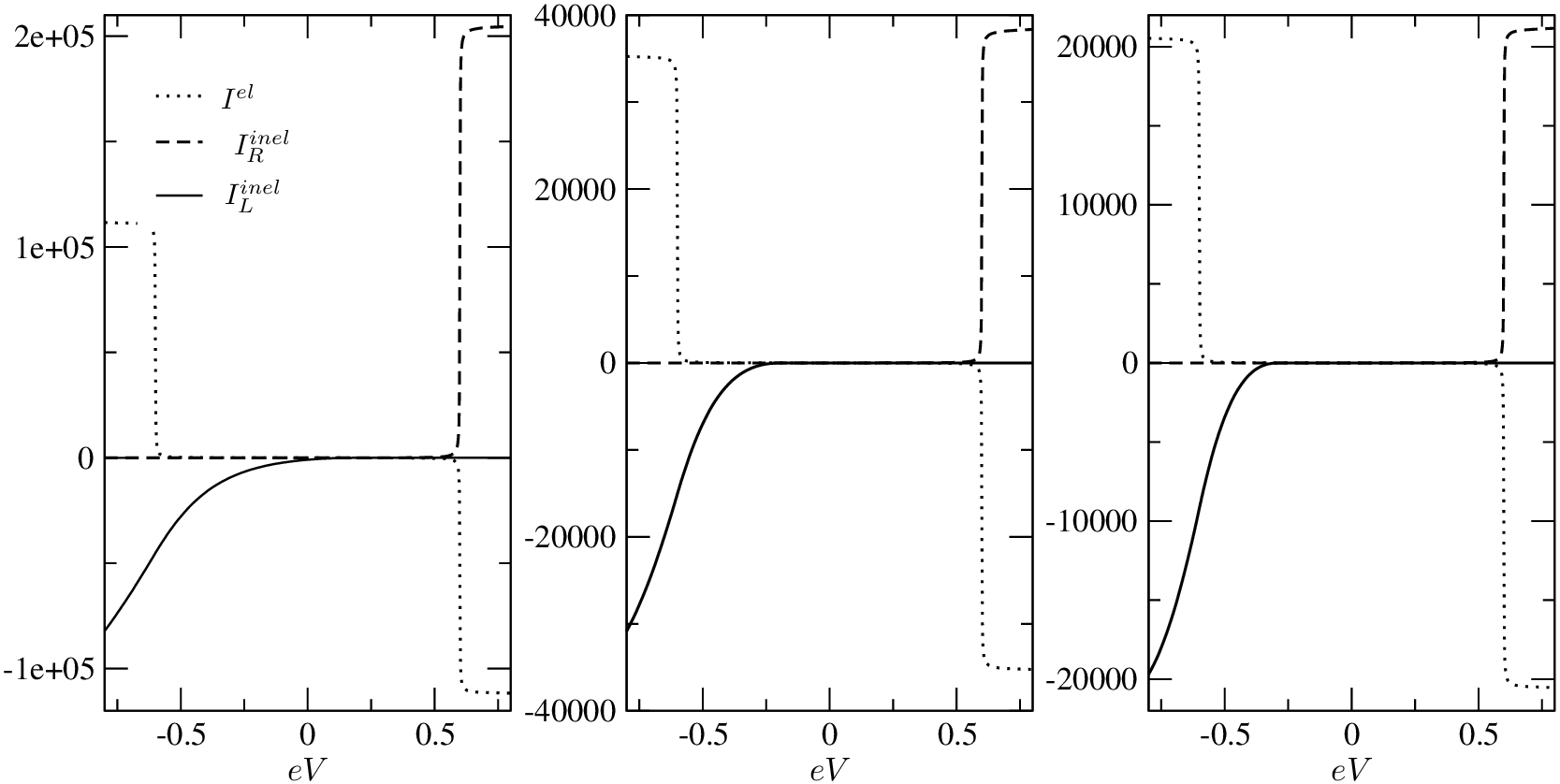}}
\caption{ Elastic and inelastic contributions to $\Delta I_{PAT}$, as a function of detector bias voltage.
The dot energy level is fixed at $\epsilon_D=0.6$. The mesoscopic device bias $eV_d$ is: 0.8 (left panel), 0.4 (center panel), and 0.3 (right panel).}
\label{fig_contribution}
\end{figure*}

We now calculate $\Delta I_{PAT}$ from Eq. (\ref{I_PAT_NDS}) with the spectral density of excess noise
 of a Quantum Point Contact, given by  Eq. (\ref{excess_noise}).
 We consider the PAT current as a function of the detector voltage $eV$
for several values of the device voltage $eV_d$, which are shown in Fig. \ref{figIPAT_eV}. We find that there
are two values of $eV$ at which $\Delta I_{PAT}$ changes drastically. First, there is a step located at $eV=\epsilon_D$
and second, a Fano-like peak appears at $eV=-\epsilon_D$. The (negative) derivative at $eV=-\epsilon_D$ seems to diverge. The high of both the peak and  the step increase in a monotonous manner as a function of the ratio of the device voltage $eV_d$ divided by the dot level $\epsilon_D$. When $eV_d$ is small, the peak is much higher than the step. Increasing $eV_d$, the peak further increases, but the step height increases faster, starting from the threshold device voltage $eV_d=\Delta-\epsilon_D$. In Fig. \ref{figIPAT_eV}, we find that for $\epsilon_D=0.4$ and $eV_d=0.8$, the peak height is still higher than the step, but with $\epsilon_D=0.6$, and $eV_d=0.6$, the step becomes higher than the peak. Here for specificity, we only consider the case
where $\epsilon_D>0$, but results for $\epsilon_D<0$ can be obtained in a similar manner, exploiting electron hole symmetry.

In order to further understand the behavior of $\Delta I_{PAT}$, we consider the different contributions of this photo-assisted current, which are shown in Fig. \ref{fig_contribution}. Specifically, we plot the elastic current renormalized by the environment, as well as the right and left inelastic currents.
We find that the elastic part is symmetric between $eV$ positive and negative. It is almost equal to zero when $|eV|<\epsilon_D$. It shows a step at $|eV|=\epsilon_D$. This can be understood from the fact that at the threshold $eV=\epsilon_D$
electrons tunnel from the normal metal lead to the superconductor predominantly by making resonant transitions
through the dot. Electrons easily tunnel to the quantum dot, in a sequential manner becoming a Cooper pair
in the superconductor.
For $eV=-\epsilon_D$, the same reasoning can be made for incoming holes, or equivalently for
electrons exiting the superconductor: a Cooper pair in the superconductor is split into
 two electrons, which tunnel to the quantum dot and then to the normal metal lead.
 Because of the energy conservation condition, it is then necessary  to have $eV< -\epsilon_D$.

Turning now to the inelastic current, we consider the contribution of 2 electrons being transfered inelastically
from the normal metal lead to the superconductor, which is called the $I^{inel}_R$ in Fig. \ref{fig_contribution}. When $eV<\epsilon_D$, electrons tunnel through the quantum dot to the superconductor by absorbing or emitting energy to environment.
However, as in the elastic case, the transfer from the normal metal to the dot is more favorable
when $eV>\epsilon_D$, which explains the presence of the step in $I^{inel}_R$ at this bias voltage.

Next, we consider the  contribution of two electrons tunneling from
the superconductor to the normal metal lead, which is called
$I^{inel}_L$ in Fig. \ref{fig_contribution}. When $eV$ is positive,
$I^{inel}_L$ is nonzero only when $eV_d>2eV$, this case corresponds
to the process of the Cooper pair absorbing energy from environment
to tunnel to the normal metal lead. If $2\epsilon_D<eV_d$, a small
step occurs (not shown) at $eV=\epsilon_D$ corresponding to the
activation of the two Cooper pair electrons on the dot: this small
feature can only be seen by zooming in the picture. When $eV$ is
negative, the absorption of energy from the environment becomes much
more favorable. Because of this, the analog of the step
corresponding to $eV=-\epsilon_D$ in the elastic current is smoothed
out, and  it saturates around $eV=-\epsilon_D$. Nevertheless,
$I^{inel}_L$ also contains contributions where electrons emit energy
to the environment. Starting from $eV<0$, the process of the Cooper
pair tunneling to the normal metal lead and emitting energy to
environment has first a small contribution to $I^{inel}_L$ but it
really becomes noticeable below $eV=-\epsilon_D$ and eventually
saturates for lower bias voltage (not shown). As the voltage of the
mesoscopic device is lowered (left to right panels of Fig.
\ref{fig_contribution}), two things occur: first, the amplitude of 
all current contributions decreases, second, the smoothing of the
$I^{inel}_L$ is reduced because the range of energy available to
absorption and emission is reduced.

In brief the sum of the contributions for emission and absorption in $I^{inel}_L$ have a tendency to
compensate the elastic current at voltages where saturation is reached.
>From the above considerations, we therefore can interpret the curve of $\Delta I_{PAT}$, and we understand when the detector circuit absorbs or emits energy from/to mesoscopic device: when $eV$ is negative, the PAT current is mainly due to absorption for $|eV|<\epsilon_D$, and emission for $|eV|>\epsilon_D$.

Note that in all our numerical results, the PAT current is in units of $2e^3R^2\gamma_1^2\gamma_2^2 T(1-T)/\pi^3\hbar^2\Delta^3 R_K$. To check the observability of these predictions we estimate: $\gamma_1=0.2\Delta$, $\gamma_2=0.02\Delta$, $\Delta=240\mu eV$, $R=0.03R_K$, $T=0.5$ (see also Ref. \onlinecite{aguado_kouwenhoven,deblock1,deblock2,recher_lossPRL}). This implies, e.g. for the PAT current in Fig. \ref{figIPAT_eV} when the detector bias is close to $\pm \epsilon_D$, $I_{PAT}\simeq 5-10 pA$ with the device bias $V_{d}=0.8\Delta/e=48\mu V$. This value is acceptable if one compares it with the value of current which has been estimated in Ref. \onlinecite{aguado_kouwenhoven}. It is also acceptable with present day detection techniques.

\section{Conclusion}

In conclusion, we have presented a new capacitive coupling scheme to study the high frequency spectral density of noise of a mesoscopic device. As in the initial proposal of Ref. \onlinecite{aguado_kouwenhoven}, the effect of the noise
originating from a mesoscopic device triggers an inelastic DC current in the detector circuit.
This inelastic contribution can be thought as a dynamical Coulomb blockade effect where the phase
fluctuations at a specific junction in the detector circuit junction are related to voltage fluctuations in same
junction. In turn, such voltage fluctuations originate from the
current fluctuations in the nearby mesoscopic device, and
both are related by a trans-impedance.
The novelty is that here, because the junction contains a superconducting element,
in the subgap regime, two electrons
need to be transferred as the elementary charge tunneling process.
In a conventional elastic tunneling situation, the two electrons injected from
(ejected in) the normal metal lead need to have exactly opposite energies
in order to combine as a Cooper pair in the superconductor.
Here, this energy conservation can be violated in a controlled manner because
a photon originating from the mesoscopic circuit can be provided to/from
the constituent electrons of the Cooper pair in the tunneling process.

We have computed the DC current in the detector circuit for two different situations.
In a first step, we considered a single NS junction, and we computed all
lowest order inelastic charge transfer processes which can be involved
in the measurement of noise: the photo-assisted transfer of
single (and pairs of) electrons (with energies within the gap)
into quasiparticle(s) above the gap,
and photo-assisted Andreev transfer of electrons as a Cooper pair
in the subgap regime.
It was shown that the latter process dominates when the source drain
bias is kept well within the gap. Close but below the gap, the absorption of
quasiparticle dominates, and we observe a crossover in the current
between the Andreev and quasiparticle contributions.
For the above processes, we demonstrated the dependence
of the DC current on the voltage bias of the mesoscopic device, chosen
here to be a quantum point contact. When this bias $eV_d$ is increased,
the overall amplitude of the spectral density and its width scale
as $V_d$, so that the phase space (the energy range) of electrons
which can contribute to the Andreev processes is magnified. We therefore
have gained an understanding about how the measurement of the DC current
can provide useful information on the noise of the mesoscopic device.
Nevertheless, one should point out that with this NS setup, it is difficult to
isolate the contribution of the photo-assisted current which involves, respectively,
the absorption and the emission of photons from the mesoscopic device.
For biases close to the chemical potential of the superconductor, both will
typically contribute to the photo-assisted current.

Next, we considered a more complex detector circuit where the normal
metal and the superconductor are separated by a quantum dot which
can only accommodate a single electron at a given time. There, the
dot acts as an additional energy filtering device, with the aim of
achieving a selection between photon emission and absorption
processes. We decided to restrict ourselves to the photo-assisted
Andreev (subgap) regime, assuming that the dot level is well within
the gap. By computing the total excess photo-assisted current as
well as its different contributions for absorption and emission, and
right and left currents, we found that for DC bias voltages
comparable to $eV= -\epsilon_D$ it is possible to make such a
distinction. The NDS detection setup could therefore provide more
information on the spectral density of noise than the NS setup, but
its diagnosis would involve the measurement of smaller currents than
the NS setup, because of the presence of two tunnel barriers instead
of one.

Note that normal lead -- quantum dot -- superconductor setups have already been investigated
theoretically\cite{kondo_effect} and experimentally recently \cite{graeber}.
In such works, the emphasis was to study how the physics of Andreev reflection affected
the Kondo anomaly in the current voltage characteristics.
In Ref. \onlinecite{graeber}, the quantum dot consisted of a carbon nanotube making
the junction between a normal metal lead and a superconductor.
Here, we did not consider the quantum dot in the Kondo regime, and we included interactions on the dot
in the Coulomb blockade regime.

A central point of this study is the fact that all contributions to the
photo-assisted current, for both the NS and NDS setups, can be cast in the same form:
\begin{eqnarray}
\Delta I_{PAT}(eV)\propto
\!\!\int\!\!d\Omega\frac{|Z(\Omega)|^2}{\Omega^2}S^+_{ex}(\pm\Omega) K_{process}(\Omega,eV,...)~.
\label{general_pat}
\end{eqnarray}
Where $K_{process}$ is a Kernel which depends on the nature  (elastic or inelastic)
as well as the mechanism (single quasi-particle or pair tunneling) of the charge transfer
process. When dealing with an elastic process, one understand that the environment
renormalizes the DC current even when no photon are exchanged between the two circuits.
In the case of inelastic tunneling only, the frequency $\Omega$ corresponds
to the total energy of the two electrons which enter (exit) the superconductor from
(to) the normal metal lead. Finally, the sign of the frequency (and thus the bound of the
integral in Eq. (\ref{general_pat}), which are left ``blank'' here) decides whether a given
contribution corresponds to the absorption or to the emission of a photon from the
mesoscopic circuit.

The present proposal has been tested using a quantum point contact as a noise source, because the spectral density of excess
noise is well characterized and because it has a simple form. It would be useful to test the present model to situations where the noise spectrum exhibits cusps or singularities. Cusps are known to occur in the high frequency (close to the gap) noise of normal superconducting junctions. Singularities in the noise are know to occur in chiral Luttinger liquid, tested in the context of the fractional quantum Hall effect\cite{frequency_fqhe}.  Such singularities or cusps should be easy to recognize in our proposed measurement of the photo-assisted current.

On general grounds, we have proposed a new mechanism which couples a normal metal -- superconductor circuit to a mesoscopic device with the goal to understand the noise
of the latter. The present setup suggest that it is plausible to extract information about high
frequency noise. High frequency noise detection now constitutes an important subfield
of nano/mesoscopic physics. New measurement setup schemes which
can be placed ``on-chip'' next to the circuit to be measured
are useful for further understanding of high frequency current moments.

\acknowledgements{We thank Richard Deblock for valuable discussions.}

\section{Appendix A}

In this Appendix, we define $\Psi_{0\leftarrow}$, $K_{2e\leftarrow}^{el}(\omega,eV,\eta)$, and $K_{2e\leftarrow}^{inel}(\Omega,eV,\eta)$ in the
\begin{eqnarray}
\Psi_{0\leftarrow}&=&\int^{-\Delta-eV}_{\Delta+eV}\!\!\!d\delta \sqrt{(\delta-eV)^2-\Delta^2}\sqrt{(\delta+eV)^2-\Delta^2}\nonumber\\
&&\times\frac{1}{\delta+i\eta}\left(\frac{1}{\delta+i\eta}-\frac{1}{\delta-i\eta}\right)~,
\label{2epsi}
\end{eqnarray}
\begin{eqnarray}
&&K_{2e\leftarrow}^{el}(\omega,eV,\eta)\nonumber\\
&=&\!\!\!\int^{-\Delta-eV}_{\Delta+eV}\!\!\!\!\!\!\!\!\!d\delta \sqrt{(\delta-eV)^2-\Delta^2}\sqrt{(\delta+eV)^2-\Delta^2}\nonumber\\
&&\times\left\{\left(2\frac{1}{\delta+i\eta}+\frac{1}{\delta+\omega+i\eta}\right)\left(\frac{1}{\delta+i\eta}-\frac{1}{\delta-i\eta}\right)\right.\nonumber\\
&&+\left.\frac{1}{\delta+i\eta}\left(\frac{1}{\delta-\omega+i\eta}-\frac{1}{\delta+\omega-i\eta}\right)\right\}~,
\label{2eKel}
\end{eqnarray}
\begin{eqnarray}
K_{2e\leftarrow}^{inel}(\Omega,eV,\eta)&=&\!\!\!\int^{\Omega-2\Delta-2eV}_{2\Delta+2eV-\Omega}\!\!\!\!\!\!d\delta\nonumber\\
&&\!\!\!\!\!\!\!\!\!\!\!\!\!\!\!\!\!\!\!\!\!\!\!\!\!\!\!\!\!\!\!\!\times\sqrt{(\Omega+\delta-2eV)^2-4\Delta^2}\sqrt{(\Omega-\delta-2eV)^2-4\Delta^2}\,\,\,\nonumber\\
&&\!\!\!\!\!\!\!\!\!\!\!\!\!\!\!\!\!\!\!\!\!\!\!\!\times \left(\frac{1}{\frac{\Omega+\delta}{2}+i\eta}-\frac{1}{\frac{\Omega-\delta}{2}-i\eta}\right)\nonumber\\
&&\!\!\!\!\!\!\!\!\!\!\!\!\!\!\!\!\!\!\!\!\!\!\!\!\!\!\!\!\!\!\!\!\!\!\!\!\!\!\!\!\!\!\!\!\!\!\!\!\!\!\!\times\left(\frac{1}{\frac{\Omega-\delta}{2}-i\eta}+\frac{1}{\frac{\Omega+\delta}{2}-i\eta}-\frac{1}{\frac{\Omega+\delta}{2}+i\eta}-\frac{1}{\frac{\Omega-\delta}{2}+i\eta}\right)\!~.
\label{2eKinel}
\end{eqnarray}

\section{Appendix B}

In this Appendix, we recall the definition of Keldysh Green's functions\cite{mahan}.
First, we define the anomalous Green's function, describing the pairing of electrons
with opposite spin in the superconductor:
$$
F_{\sigma}(q,t-t^\prime)\equiv -\langle T_K c_{-q,-\sigma}(t)c_{q,\sigma}(t^\prime)\rangle~,
$$
$$
F_{\sigma}^\dagger(q,t-t^\prime)\equiv\langle T_K c^\dagger_{q,\sigma}(t)c_{-q,-\sigma}^\dagger(t^\prime)\rangle~.
$$
If both $t$ and $t^\prime$ are in the upper branch, and $t>t^\prime$ or both $t$ and $t^\prime$ are in the lower branch, and $t^\prime>t$ then
$F_{\sigma}(t_+-t^\prime_+)=F^\dagger_{\sigma}(t_--t_-^\prime)=\text{sgn}(\sigma) u_q v_q e^{-iE_q (t-t^\prime)}$.
These Green's functions enter the description of the Andreev process.

If we consider the single quasiparticle tunneling in the superconductor, we use the conventional definition
of the Green's function as for normal metals.


Secondly, we define the Green's function of the one level QD:
$G_{D\sigma}(t-t^\prime)\equiv\langle T_K c_{D\sigma}(t)c^{+}_{D\sigma}(t^\prime)\rangle$.

Simplifications occur because the quantum dot has a singly occupied level with energy $\epsilon_D$, the first electron is transferred to the lead before the second hops on the quantum dot so that in our work, we only consider the QD Green's function where both time quantities $t$ and $t^\prime$ are in the upper or lower branch, and the Green's function values only when $t>t^\prime$ if $t$, $t^\prime$ in the upper branch $G^{t}_{D\sigma}(t-t^\prime)=e^{-i\epsilon_D (t-t^\prime)}$ and $t^\prime>t$ if $t$, $t^\prime$ in the lower branch, then $G^{\tilde{t}}_{D\sigma}(t-t^\prime)=e^{-i\epsilon_D (t-t^\prime)}$.

The Green's function in the normal metal lead reads:
$G_{L\sigma}(k,t-t^\prime)\equiv \langle T_K c_{k\sigma}(t)c^\dagger_{k\sigma}(t^\prime)\rangle$.

In our work, we consider the cases where two electrons tunneling from/to normal metal lead, so that we only consider normal metal Green's function where $t$ and $t^\prime$ are in the different branches. For the case electrons tunneling from the superconductor to the normal metal, we use the greater Green's function $G^>_{L\sigma}(k,t-t^\prime)=e^{-i(\epsilon_k-\mu_L) (t-t^\prime)}$ with $\epsilon_k >\mu_L$. For the case electrons tunneling from the normal metal to the superconductor, we use the lesser Green's function $G^<_{L\sigma}(k,t-t^\prime)=-\langle c^\dagger_{k\sigma}(t^\prime)c_{k\sigma}(t)\rangle =-e^{-i(\epsilon_k-\mu_L) (t-t^\prime)}$ with $\epsilon_k\le\mu_L$.

\section{Appendix C}

In this part, we present the denominator products  which appear in the tunneling current through the NS junction as a Cooper pair. Such denominators come from the energy denominators of the transition operator T.
\begin{equation}
\!\!(D^0_\leftarrow)^{-1}\!\!=\!\frac{1}{E^\prime+\epsilon+i\eta}\left(\frac{1}{E-\epsilon-i\eta}+\frac{1}{E+\epsilon-i\eta}\right)~,
\label{denominatorNS1}
\end{equation}
\begin{eqnarray}
(D^{el}_\leftarrow)^{-1}\!&=&\!\frac{1}{E^\prime+\epsilon+\omega+i\eta}\left(\frac{1}{E-\epsilon-i\eta}+\frac{1}{E+\epsilon-i\eta}\right)\,\,\,\nonumber\\
&&\!\!\!\!\!\!\!\!\!\!\!\!\!\!\!\!\!\!\!\!\!\!\!\!\!\!\!\!\!\!\!+\frac{1}{E^\prime+\epsilon+i\eta}\left(\frac{1}{E-\epsilon+\omega-i\eta}+\frac{1}{E+\epsilon+\omega-i\eta}\right)\!~,
\label{denominatorNS2}
\end{eqnarray}
\begin{eqnarray}
(D^{inel}_\leftarrow)^{-1}\!\!&=&\!\!\left(\frac{1}{E^\prime-\epsilon^\prime+i\eta}+\frac{1}{E^\prime+\epsilon+i\eta}\right)\nonumber\\
&&\times\left(\frac{1}{E+\epsilon^\prime-i\eta}+\frac{1}{E+\epsilon-i\eta}\right.\nonumber\\
&&+\left.\frac{1}{E-\epsilon-i\eta}+\frac{1}{E-\epsilon^\prime-i\eta}\right)\!~.
\label{denominatorNS3}
\end{eqnarray}

Change variables in $I^{inel}_\rightarrow$, $I^{inel}_\leftarrow$ as
\[ \left\{
\begin{array}{l l}
    \Omega=\epsilon+\epsilon^\prime~,\\
       \delta=\epsilon-\epsilon^\prime~.\end{array} \right. \]
We define
\begin{eqnarray}
\chi(x,\eta)&\equiv&\int_{\Delta}^{\infty}\!\!\!\! dE\frac{1}{\sqrt{E^2-\Delta^2}}\frac{1}{E-x-i\eta}\nonumber\\
&=&\frac{\pi+2\arcsin(\frac{x+i\eta}{\Delta})}{2\sqrt{\Delta^2-(x+i\eta)^2}}~,
\end{eqnarray}
then we define the weight functions as
\begin{eqnarray}
&&K^{el}(\omega,eV,\eta)\nonumber\\
&=&\!\!\!\int_{-eV}^{eV}\!\!\!\!\!\!d\epsilon \left\{\left[2\chi(-\epsilon,-\eta)+\chi(-\epsilon-\omega,-\eta)\right]\left[\chi(\epsilon,\eta)+\chi(-\epsilon,\eta)\right]\right.\nonumber\\
&&+\left.\chi(-\epsilon,-\eta)\left[\chi(\epsilon-\omega,\eta)+\chi(-\epsilon-\omega,\eta)\right]\right\}~,
\label{kernelNSel}
\end{eqnarray}
\begin{eqnarray}
K^{inel}(\Omega,eV,\eta)\!&=&\!\!\!\int_{2eV-\Omega}^{\Omega-2eV}\!\!\!\!\!\!\!\!\!d\delta \left[\chi(\frac{\Omega-\delta}{2},-\eta)+\chi(-\frac{\Omega+\delta}{2},-\eta)\right]\nonumber\\
&&\!\!\!\!\!\!\!\!\!\times\left[\chi(-\frac{\Omega-\delta}{2},\eta)+\chi(-\frac{\Omega+\delta}{2},\eta)\right.\nonumber\\
&&\left.+\chi(\frac{\Omega+\delta}{2},\eta)+\chi(\frac{\Omega-\delta}{2},\eta)\right]~.
\label{kernelNSinel}
\end{eqnarray}

\section{Appendix D}

In this Appendix, we present the denominator products  which appear in the tunneling current through the NDS junction. 

$D^0$ is the original denominator which is not affected by the environment
\begin{widetext}
\begin{eqnarray}
(D^0_\leftarrow)^{-1}&=&\frac{1}{(E^\prime+\epsilon+i\eta)(E^\prime+\epsilon_D+i\eta)(\epsilon+\epsilon_D+i\eta)(E+\epsilon_D-i\eta)}\nonumber\\
&&\times\left[\frac{1}{(-\epsilon+\epsilon_D-i\eta)(E-\epsilon-i\eta)}+\frac{1}{(\epsilon+\epsilon_D-i\eta)(E+\epsilon-i\eta)}\right]~,
\label{denominatorNDS1}
\end{eqnarray}
and $D^{el}_\leftarrow$ is the denominator product affected by the environment
\begin{eqnarray}
(D^{el}_\leftarrow)^{-1}&=&\frac{1}{(E^\prime+\epsilon+\omega+i\eta)(E^\prime+\epsilon_D+\omega+i\eta)(\epsilon+\epsilon_D+i\eta)(E+\epsilon_D-i\eta)}\nonumber\\
&&\times\left[\frac{1}{(-\epsilon+\epsilon_D-i\eta)(E-\epsilon-i\eta)}+\frac{1}{(\epsilon+\epsilon_D-i\eta)(E+\epsilon-i\eta)}\right]\nonumber\\
&&+\frac{1}{(E^\prime+\epsilon+i\eta)(E^\prime+\epsilon_D+i\eta)(\epsilon+\epsilon_D+i\eta)(E+\epsilon_D+\omega-i\eta)}\nonumber\\
&&\times\left[\frac{1}{(-\epsilon+\epsilon_D-i\eta)(E-\epsilon+\omega-i\eta)}+\frac{1}{(\epsilon+\epsilon_D-i\eta)(E+\epsilon+\omega-i\eta)}\right]~,
\label{denominatorNDS2}
\end{eqnarray}
where $D^{inel}$ is the denominator product attributed to the inelastic current (affected by environment) and it is defined as
\begin{eqnarray}
(D^{inel}_\leftarrow)^{-1}&=&\left[\frac{1}{(E^\prime+\epsilon_D-\epsilon-\epsilon^\prime+i\eta)(E^\prime-\epsilon^\prime+i\eta)}+\frac{1}{(E^\prime+\epsilon_D+i\eta)(E^\prime+\epsilon+i\eta)}\right]\nonumber\\
&&\!\!\!\!\!\!\!\!\!\!\!\!\times\left\{\frac{1}{(\epsilon_D-\epsilon^\prime+i\eta)(\epsilon_D-\epsilon-i\eta)}\left[\frac{1}{(E+\epsilon_D-\epsilon-\epsilon^\prime-i\eta)(E-\epsilon-i\eta)}+\frac{1}{(E+\epsilon_D-i\eta)(E+\epsilon^\prime-i\eta)}\right]\right.\nonumber\\
&&\!\!\!\!\!\!\!\!\!\!\!\!\!\!\!\!\!\!+\left.\frac{1}{(\epsilon_D-\epsilon^\prime+i\eta)(\epsilon_D-\epsilon^\prime-i\eta)}\left[\frac{1}{(E+\epsilon_D-\epsilon-\epsilon^\prime-i\eta)(E-\epsilon^\prime-i\eta)}+\frac{1}{(E+\epsilon_D-i\eta)(E+\epsilon-i\eta)}\right]\right\}~.
\label{denominatorNDS3}
\end{eqnarray}

We define
\begin{equation}
\Pi(x_1,x_2,\eta)\!=\!\!\!\int_{\Delta}^{\infty}\!\!\!\!\!\!\!dE\frac{1}{\sqrt{E^2-\Delta^2}}\frac{1}{(E-x_1-i\eta)(E-x_2-i\eta)}~.
\end{equation}
If $x_1\neq x_2$ then
$$
\Pi(x_1,x_2,\eta)=\frac{1}{2(x_1-x_2)}\left( \frac{\pi+2\arcsin(\frac{x_1+i\eta}{\Delta})}{\sqrt{\Delta^2-(x_1+i\eta)^2}}-\frac{\pi+2\arcsin(\frac{x_2+i\eta}{\Delta})}{\sqrt{\Delta^2-(x_2+i\eta)^2}} \right)~,
$$
else
$$
\Pi(x_1,x_2,\eta)=\frac{(x_1+i\eta)(\pi+2\arcsin(\frac{x_1+i\eta}{\Delta}))}{2(\Delta^2-(x_1+i\eta)^2)^{3/2}}+\frac{1}{\Delta^2-(x_1+i\eta)^2}~.
$$
Then we define
\begin{eqnarray}
\Psi^0_\leftarrow(\epsilon,\epsilon_D,\eta)&=&\int_{\Delta}^{\infty}\!\!\! dE\int_{\Delta}^{\infty}\!\!\! dE^\prime\frac{1}{\sqrt{E^2-\Delta^2}\sqrt{{E^\prime}^2-\Delta^2}}\frac{1}{D^0_\leftarrow}\nonumber\\
&=&\Pi(-\epsilon,-\epsilon_D,-\eta)\left\{\frac{\Pi(\epsilon,-\epsilon_D,\eta)}{(\epsilon+\epsilon_D+i\eta)(-\epsilon+\epsilon_D-i\eta)}+\frac{\Pi(-\epsilon,-\epsilon_D,\eta)}{(\epsilon+\epsilon_D+i\eta)(\epsilon+\epsilon_D-i\eta)}\right\}~,
\end{eqnarray}
\begin{eqnarray}
\Psi^{el}_\leftarrow(\epsilon,\epsilon_D,\omega,\eta)&=&\int_{\Delta}^{\infty}\!\!\! dE\int_{\Delta}^{\infty}\!\!\! dE^\prime\frac{1}{\sqrt{E^2-\Delta^2}\sqrt{{E^\prime}^2-\Delta^2}}\frac{1}{D^{el}_\leftarrow}\nonumber\\
&=&\frac{\Pi(-\epsilon-\omega,-\epsilon_D-\omega,-\eta)\Pi(\epsilon,-\epsilon_D,\eta)+\Pi(\epsilon-\omega,-\epsilon_D-\omega,\eta)\Pi(-\epsilon,-\epsilon_D,-\eta)}{(\epsilon+\epsilon_D+i\eta)(-\epsilon+\epsilon_D-i\eta)}\nonumber\\
&&+\frac{\Pi(-\epsilon-\omega,-\epsilon_D-\omega,-\eta)\Pi(-\epsilon,-\epsilon_D,\eta)+\Pi(-\epsilon-\omega,-\epsilon_D-\omega,\eta)\Pi(-\epsilon,-\epsilon_D,-\eta)}{(\epsilon+\epsilon_D+i\eta)(\epsilon+\epsilon_D-i\eta)}~,
\end{eqnarray}
\begin{eqnarray}
\Psi^{inel}_\leftarrow(\epsilon,\epsilon^\prime,\epsilon_D,\eta)&=&\int_{\Delta}^{\infty}\!\!\! dE\int_{\Delta}^{\infty}\!\!\! dE^\prime\frac{1}{\sqrt{E^2-\Delta^2}\sqrt{{E^\prime}^2-\Delta^2}}\frac{1}{D^{inel}_\leftarrow}\nonumber\\
&=&\left(\Pi(\epsilon+\epsilon^\prime-\epsilon_D,\epsilon^\prime,-\eta)+\Pi(-\epsilon_D,-\epsilon,-\eta)\right)\nonumber\\
&&\times\left\{\frac{\Pi(\epsilon+\epsilon^\prime-\epsilon_D,\epsilon,\eta)+\Pi(-\epsilon_D,-\epsilon^\prime,\eta)}{(\epsilon_D-\epsilon^\prime+i\eta)(\epsilon_D-\epsilon-i\eta)}
+\frac{\Pi(\epsilon+\epsilon^\prime-\epsilon_D,\epsilon^\prime,\eta)+\Pi(-\epsilon_D,-\epsilon,\eta)}{(\epsilon_D-\epsilon^\prime+i\eta)(\epsilon_D-\epsilon^\prime-i\eta)}\right\}~.
\end{eqnarray}
\end{widetext}
Since $D^{0}_\rightarrow(\epsilon)=D^{0}_\leftarrow(-\epsilon)$, $D^{el}_\rightarrow(\epsilon)=D^{el}_\leftarrow(-\epsilon)$, so $\Psi^{0}_\rightarrow(\epsilon)=\Psi^{0}_\leftarrow(-\epsilon)$, $\Psi^{el}_\rightarrow(\epsilon)=\Psi^{el}_\leftarrow(-\epsilon)$. But in fact, if we change the name of variable $\epsilon \rightarrow \epsilon^\prime$ then change variable $\epsilon^\prime=-\epsilon$, we will obtain the same formula for both cases $eV>0$ and $eV<0$. Since $\epsilon, \epsilon^\prime$ are independently equivalent, so that it is evidently $\Psi^{inel}_\rightarrow(\epsilon,\epsilon^\prime,\epsilon_D,\eta)=\Psi^{inel}_\leftarrow(\epsilon,\epsilon^\prime,\epsilon_D,\eta)$. Hereafter, we neglect the $\leftarrow$ or $\rightarrow$ index in these functions.

If $eV>0$, the elastic current contributions in $I_\rightarrow$ exist but the elastic current contributions in $I_\leftarrow$ vanish (in contrast to the case of $eV<0$).

We change variables in inelastic contributions as

\[ \left\{
\begin{array}{l l}
    \Omega=\epsilon+\epsilon^\prime~,\\
       \delta=\epsilon-\epsilon^\prime~,\end{array} \right. \]
and define
\begin{equation}
K^{el}_{NDS}(\omega,eV,\epsilon_D,\eta)\!=\!\!\int_{-eV}^{eV}\!\!\!\! d\epsilon \Psi^{el.tot}(\epsilon,\epsilon_D,\eta)~,
\end{equation}

\vspace{-0.8cm}

\begin{eqnarray}
&&\!\!\!\!\!\!\!\!\!\!\!\!\!\!\!\!\!\!\!\!K^{inel}_{NDS}(\Omega,eV,\epsilon_D,\eta)\nonumber\\
&=&\!\!\int_{2eV-\Omega}^{\Omega-2eV}\!\!\!\!\!\!\!\!d\delta \Psi^{inel}(\frac{\Omega+\delta}{2},\frac{\Omega-\delta}{2},\epsilon_D,\eta)~,
\end{eqnarray}
with $\Psi^{el.tot}(\epsilon,\epsilon_D,\omega,\eta)=2\Psi^{0}(\epsilon,\epsilon_D,\eta)+\Psi^{el}(\epsilon,\epsilon_D,\omega,\eta)$.


\begin{thebibliography}{99}
\bibitem{notice} In explicit calculations we set $\hbar=1$ but restore $\hbar$ in the fundamental quantities and in the numerical estimate of the PAT current.
\end{thebibliography}
\end{document}